\documentclass{article}
\pdfoutput=1 
\usepackage{url}
\usepackage{amsmath,amssymb}
\usepackage{graphicx}
\usepackage{url}
\usepackage{listings} 
\usepackage{color} 
\usepackage{textcomp} 
\usepackage{color}
\usepackage[all]{xy}

\usepackage{authblk}

\usepackage{pdflscape}

\newcommand \defn {\mathrel{\triangleq}}

\newcommand \E {\mathbb{E}}
\newcommand \thr {\tau}
\newcommand \err {\varepsilon}
\newcommand \noise {\omega}
\newcommand \loss {L}
\newcommand \pa {p_A}
\newcommand \pt {p_T}
\newcommand \LA {\ell_A}
\newcommand \LT {\ell_T}
\newcommand \AND {\wedge}

\newtheorem{theorem}{Theorem}

\newtheorem{lemma}{Lemma}
\newtheorem{corollary}{Corollary}
\newtheorem{definition}{Definition}
\newtheorem{assumption}{Assumption}
\newtheorem{remark}{Remark}

\newenvironment{proof}[1][Proof]{\begin{trivlist}
\item[\hskip \labelsep {\bfseries #1}]}{\end{trivlist}}

\newcommand{\qed}{\nobreak \ifvmode \relax \else
      \ifdim\lastskip<1.5em \hskip-\lastskip
      \hskip1.5em plus0em minus0.5em \fi \nobreak
      \vrule height0.75em width0.5em depth0.25em\fi}

\definecolor{gray}{gray}{0.5}
\definecolor{key}{rgb}{0,0.5,0}
\lstnewenvironment{python}[1][]{
\lstset{
language=python,
basicstyle=\ttfamily\small,
otherkeywords={1, 2, 3, 4, 5, 6, 7, 8 ,9 , 0, -, =, +, [, ], (, ), \{, \}, :, *, !},
keywordstyle=\color{blue},
stringstyle=\color{red},
showstringspaces=false,
emph={class, pass, in, for, while, if, is, elif, else, not, and, or,
def, print, exec, break, continue, return},
emphstyle=\color{black}\bfseries,
emph={[2]True, False, None, self},
emphstyle=[2]\color{key},
emph={[3]from, import, as},
emphstyle=[3]\color{blue},
upquote=true,
morecomment=[s]{"""}{"""},
commentstyle=\color{gray}\slshape,
framexleftmargin=1mm, framextopmargin=1mm, frame=shadowbox,
rulesepcolor=\color{blue},#1
}}{}

\newcommand{\protheader}[5]{\vspace{1.5ex} 
\begin{minipage}[t]{0.125\textwidth}{}\end{minipage} \hfill
\begin{minipage}[t]{0.3\textwidth}{\fbox{#1}} \\[1.5ex]
#2 \end{minipage} \hfill
\begin{minipage}[t]{0.15\textwidth}{\begin{center}Channel \\[1.5ex] #3 \end{center}}\end{minipage} \hfill
\begin{minipage}[t]{0.3\textwidth}{\begin{flushright} \fbox{#4} \\[1.5ex] #5 \end{flushright}}
\end{minipage} \hfill
\begin{minipage}[t]{0.125\textwidth}{}\end{minipage}
\begin{minipage}[t]{\textwidth}
\setlength{\unitlength}{\textwidth}
\begin{picture}(1,0.06)
\put(0,0.03){\line(1,0){1}} \put(0.08,0.05){\vector(0,-1){0.04}} 
\put(0.92,0.05){\vector(0,-1){0.04}} 
\end{picture}
\end{minipage}
}

\newcommand{\protline}[3]{\noindent
\begin{minipage}[t]{0.01\textwidth}{}\end{minipage} \hfill
\begin{minipage}[t]{0.41\textwidth}{#1}\end{minipage} \hfill
\begin{minipage}[t]{0.2\textwidth}{#2}\end{minipage} \hfill
\begin{minipage}[t]{0.35\textwidth}{\begin{flushright} #3
\end{flushright}} \end{minipage} \hfill
\begin{minipage}[t]{0.01\textwidth}{}\end{minipage}
\\[1.5ex]
}

\newcommand{\protrightarrow}[1]{
\setlength{\unitlength}{\textwidth}
\begin{picture}(1,0.1)
\put(0,-0.03){\vector(1,0){1}} 
\put(0,0){\begin{minipage}[t]{\textwidth}
\begin{center} #1 \end{center}
\end{minipage}}
\end{picture}
}

\newcommand{\protleftarrow}[1]{
\setlength{\unitlength}{\textwidth}
\begin{picture}(1,0.1)
\put(1,-0.03){\vector(-1,0){1}}
\put(0,0){\begin{minipage}[t]{\textwidth}
\begin{center} #1 \end{center}
\end{minipage}}
\end{picture}
}

\newcommand{\protocol}[1]{\noindent \fbox{\parbox{\textwidth}{#1}}}

\newcommand{\leer}{\hspace{1em}}

\begin{document}
\title{Shedding Light on RFID Distance Bounding Protocols and Terrorist Fraud Attacks}
\author{Pedro Peris-Lopez\footnote{Corresponding Author: Delft University of Technology (TU-Delft),
Faculty of Electrical Engineering, Mathematics, and Computer Science (EEMCS),
Security Lab. P.O. Box 5031
2600 GA, Delft, The Netherlands; E-mail: P.PerisLopez@tudelft.nl.}, Julio C. Hernandez-Castro, Christos Dimitrakakis, Aikaterini Mitrokotsa, Juan M. E. Tapiador}

\maketitle
\begin{abstract}
The vast majority of RFID authentication protocols assume the proximity between readers and tags due to the limited range of the radio channel. However, in real scenarios an intruder can be located between the prover (tag) and the verifier (reader) and trick this last one into thinking that the prover is in close proximity. This attack is generally known as a relay attack in which scope distance fraud, mafia fraud and terrorist attacks are included.  Distance bounding protocols represent a promising countermeasure to hinder relay attacks. Several protocols have been proposed during the last years but vulnerabilities of major or minor relevance have been identified in most of them.  In 2008, Kim et al. \cite{KimAKSP-2008-icisc} proposed a new distance bounding protocol with the objective of being the best in terms of security, privacy, tag computational overhead and fault tolerance. In this paper, we analyze this protocol and we present a passive full disclosure attack, which allows an adversary to discover the long-term secret key of the tag.  The presented attack is very relevant, since no security objectives are met in Kim et al.'s protocol. Then, design guidelines are introduced with the aim of facilitating protocol designers the stimulating task of designing secure and efficient schemes against relay attacks. Finally a new protocol, named Hitomi and inspired by \cite{KimAKSP-2008-icisc}, is designed conforming the guidelines proposed previously.

\end{abstract}

\section{Introduction}
Radio-Frequency Identification (RFID) is one of the most promising technologies in identifying items (i.e. persons, animals or products) with high accuracy and it overcomes all the other relevant technologies (i.e. barcodes) \cite{Jues-survey-2005}. While barcodes facilitate the identification process of the brand and model of a product, RFID tags offer the possibility to distinguish products of the same kind (i.e. unequivocal identification of labeled items).  Specifically, an RFID system is composed of three main components: tags, readers and a back-end database. The readers
(transceivers) interrogate tags (transponders) to access the
information stored in their memory. Afterwards, they pass this
acquired information to a back-end database which employs it as a search
index to allocate all the information associated with the
target tag. Readers and tags use a radio channel for
communication, which is commonly assumed to be insecure. On
the other hand, readers and the back-end database have sufficient power to provide full cryptographic security.

RFID technology  present a lot of advantages and broad applicability. However, the massive adoption of this technology is
delayed due to its associated security threats \cite{Jues-survey-2005,MitrokotsaRT-2008-iwrt}. In this paper, we focus on relay attacks, a type of attack which has been gaining attention recently. For example,  Hlavac and Rosa \cite{cryptoeprint:2007:244} notice how proximity cards conforming to ISO/IEC 14443 can be abused by a relay attack exploiting the extended timeouts in the communication.

\section{Related Work}
Generally, a relay attack is a kind of a man-in-the-middle attack, in which an attacker relays messages from an authentic tag to a legitimate reader. If the attack is successful, the adversary tricks a valid reader into believing that it is communicating with a valid tag and that this tag is within a particular physical distance. As an alternative, the signal strength could be used as a measure for the detection of relay attacks. Nevertheless, this approach is not effective when the adversary is more sophisticated and transmits with much more power than expected. Distance bounding protocols were  first introduced by Brands and Chaum \cite{brands94} to preclude distance fraud and mafia fraud attacks. The authors proposed a mechanism to infer an upper bound of the distance between the verifier and the prover by measuring the round trip delay during a rapid challenge-response bit exchange of $n$ rounds.  Around fifteen years later, Hancke and Kuhn \cite{hancke05} proposed a distance bounding protocol in the context of RFID technology which may be considered a seminal paper in this research area. Later, Munilla and Peinado \cite{munilla2} proposed a protocol inspired by \cite{hancke05} in which the success probability of an adversary to accomplish a mafia fraud attack is reduced. However, the feasibility of this scheme is questionable since it requires three physical states \{0, 1, void\}.  Singelee and Preneel \cite{singeleep07} proposed a distance bounding protocol which uses an error correction code to facilitate the corrections of errors (in noisy channels) during the rapid bit exchange. Nevertheless, this scheme's security and implementation cost on RFID tags is questioned in~\cite{MunillaP-2010-elseviercc}. Finally, the above mentioned protocols do not address {\em terrorist fraud} attacks, which can be considered an extension of mafia fraud attacks.  In 2007, Tu and Piramuthu \cite{TuP-2007-rfidtechnology} addressed both terrorist and mafia fraud attacks and proposed an enhancement scheme. The authors used ideas previously presented in \cite{reid2007} to prevent terrorist attacks.  Nevertheless, Kim et al. \cite{KimAKSP-2008-icisc} noted that Tu's and Piramuthu's protocol is vulnerable to a simple active attack and proposed a new protocol attempting to correct the defenses of all its predecessors.\\

\textbf{Our contribution.} In this paper, we analyze the protocol proposed by Kim et al. \cite{KimAKSP-2008-icisc}, which may currently be considered the most secure and efficient distance bounding protocol in the class of protocols that include a final signature. However, we show that their protocol presents a vulnerability which renders it insecure to a passive attack. We need to note here that passive attacks are much less exigent than active attacks (e.g. mafia and terrorist fraud attacks) since the attacker only has to eavesdrop the messages transmitted on the channel. As a consequence of the passive attack presented below,  the attacker can acquire the full long-term secret key of the tag. We also provide some guidelines for designing secure and efficient distance bounding protocols that are resistant to relay attacks and passive eavesdroppers. Finally, we introduce -- and provide a security and performance analysis --  the Hitomi RFID distance bounding protocol which complies with the proposed guidelines and is suitable for constrained devices.

\section{Relay Attacks}

\begin{figure}
\centering
\includegraphics[width=8cm]{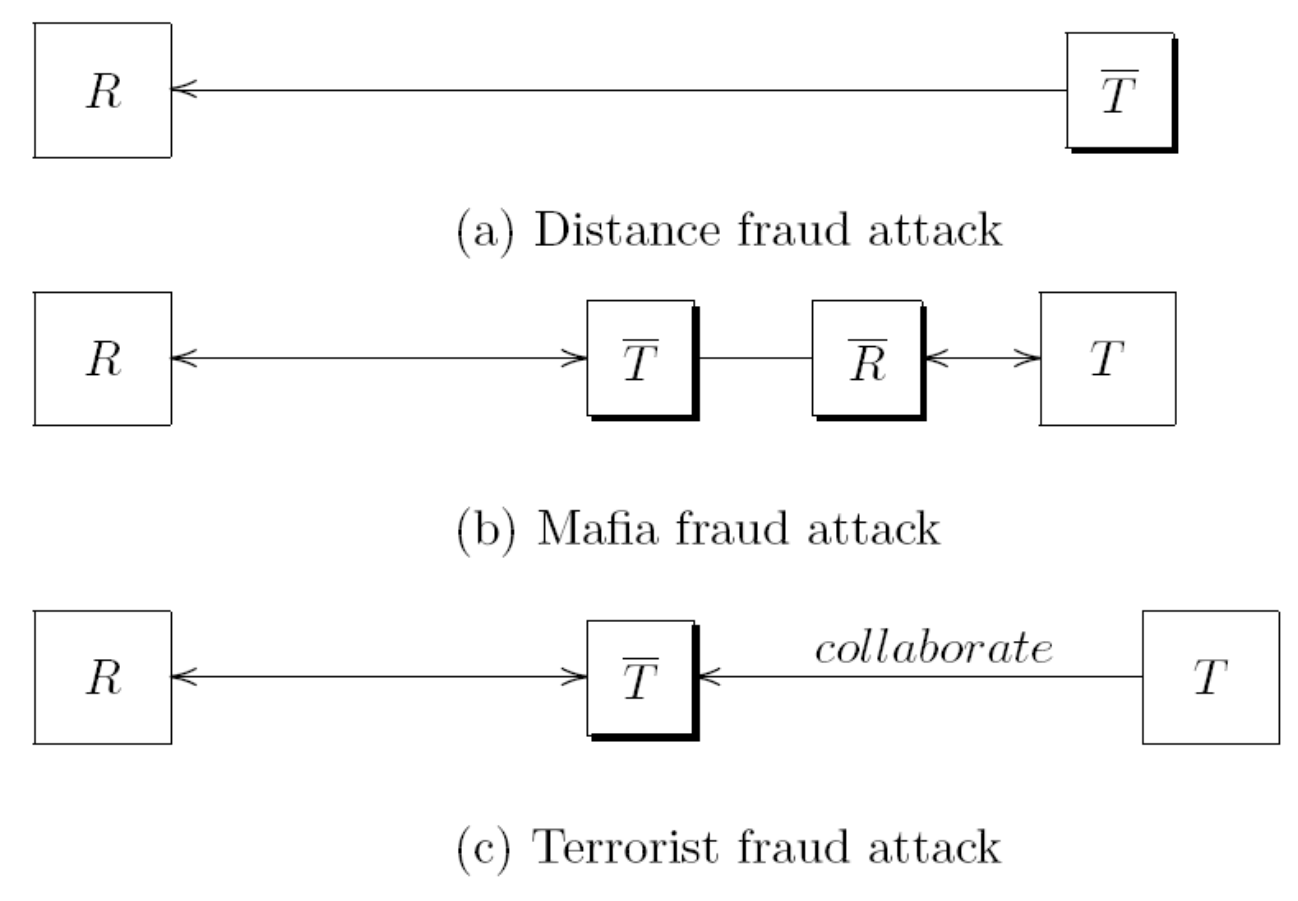}\\
\caption{Distance, Mafia and Terrorist Fraud Attack }
\label{fig::relay}
\end{figure}

When designing an RFID distance bounding protocol three real-time frauds \cite{Bussard-2004-thesis,BussardB05} need to be addressed: 1) distance fraud attacks; 2) mafia fraud attacks; 3) terrorist fraud attacks (Fig. \ref{fig::relay}).

\begin{definition}
In distance fraud, two entities are involved: the first (the honest reader $R$) is not aware of the attack performed by the second party (the fraudulent tag $\overline{T}$).  The fraud enables $\overline{T}$ to convince $R$ of a wrong statement related to its physical distance to $R$.
\end{definition}

\begin{definition}
In mafia fraud, three entities are involved: the two first (honest tag $T$ and reader $R$) are not aware of the attack performed by the third party (intruder $I$). The fraud enables $I$ to convince $R$ of an assertion related to the private key of $T$.
\end{definition}
The mafia fraud was first described by Desmedt \cite{desmedt}. In this fraud, the intruder is modeled as a couple $\{\overline{T}, \overline{R}\}$, where $\overline{T}$ is a dishonest tag interacting with the honest reader $R$ and  $\overline{R}$ is a dishonest reader interacting with the honest tag $T$. With the help of $\overline{R}$, $R$ is convinced by $\overline{T}$ of an assertion related to the private key of $T$. Specifically, the assertion consists on the fact that the tag $\overline{T}$ is within a particular physical distance.

\begin{definition}
In terrorist fraud, three entities are involved: the first (the reader $R$) is not aware of the attack performed by the two collaborating parties (the dishonest tag ${T}$ and the intruder or terrorist tag $\overline{T}$). The fraud enables $\overline{T}$ to convince $R$ of an assertion related to the private key of $T$.
\end{definition}

This attack can be viewed as an extension of the mafia fraud attack. In this fraud, the dishonest tag $T$ collaborates with the terrorist tag $\overline{T}$. The dishonest tag $T$ uses $\overline{T}$ to convince $R$ that it lies in close proximity, while in fact it does not. Nevertheless, the long-term secret key of $T$ is not revealed to the terrorist tag $\overline{T}$.

Apart from distance bounding protocols, constrained channel, context sharing, isolation, unforgeable channel, time of flight are general techniques which offer complete or partially resistance to relay attacks. In fact, only the isolation technique offers protection against distance, mafia and terrorist fraud attacks.  We urge the reader to consult \cite{BussardB05} for a complete description of these techniques.

\section{Terrorist Fraud Attacks}

This section briefly describes a selection of distance bounding protocols that aim to guard against terrorist fraud attacks. The reader is urged to consult each cited reference for a detailed description of each approach.

In \cite{BussardB05}, Bussard and Bagga addressed the fraud where a malicious prover and an intruder collaborate to cheat a verifier.  A secret sharing strategy is proposed to combat terrorist fraud attacks. More precisely, the prover picks a random one-time key $a$ and encrypts its private long-term key $x$ according to a publicly known symmetric encryption algorithm $\{E_a(x)\}$. The prover then splits his permanent secret key into two shares by computing $Z^{0} := a$ and $Z^1 := E_a(x)$.  Apart from the distance-bounding stage, the whole scheme is completed by a bit commitment scheme and a proof of knowledge stage based on public cryptography. The resources (i.e. computation, storage, power consumption, etc.) needed to support on-chip these two stages render this approach useless for constrained devices such as low-cost RFID tags.

Reid et al. \cite{reid2007} replaced asymmetric cryptography by symmetric cryptography in order to facilitate implementation on devices with limited resources (i.e. sensor networks, RFID tags, etc.). The prover computes a session key $\{a := f_x(ID_A, ID_B, r_A, r_B)\}$, where $f$ denotes a keyed hash function. Finally, the prover splits his permanent secret key into two shares by computing $Z^{0} := a$ and $Z^1 := a \oplus x$.

Tu and Piramuthu \cite{TuP-2007-rfidtechnology} proposed a new protocol arguing that in \cite{reid2007} the identities of the prover and the verifier are transmitted in clear and thus, can be easily traced.  Nevertheless, Tu and Piramuthu do not take into consideration that Reid et al.'s protocol is not proposed in the context of RFID where anonymity and untraceability is a security objective.  On the other hand, they claim -- as in  \cite{KimAKSP-2008-icisc} -- that the probability with which a mafia fraud attack can occur is bounded by $(7/8)^n$ \cite{piramuthu07}. This argumentation is completely incorrect because Tu and Piramuthu conclude this based on the fact that $f_x(ID_A, ID_B, r_A, r_B)$ and $f_x(ID_A, ID_B, r_A', r_B)$ only differ in $1/4$ of the bits given that the only argument that changes is $r_A$, while the rest remain constant. However,  at least half of the bits are changing since $f$ is a keyed hash function (e.g. CBC-MAC or HMAC) as suggested in \cite{reid2007}.
\begin{theorem}\label{th::Reid}
In Reid et al.'s protocol, the probability that a mafia fraud attack can occur is bounded by $(\frac{3}{4})^n$ \cite{Mitro1kotsaDPH10}.
\end{theorem}

For a detailed proof of this theorem, the reader is urged to consult \cite{Mitro1kotsaDPH10}. Although we recommend reading the original paper, we include a sketch proof for completeness.

\begin{proof}[Sketch Proof]
An adversary could slightly accelerate the clock signal provided to the tag and transmit an anticipated challenge $c'_i$ before the reader sends its challenge $c_i$ to the tag. In half of the times, these values fit in, that is $c'_i = c_i$, and therefore the adversary will have in advance the correct answer $r_i$ to the reader. In the other half of the cases, the adversary can transmit a guessed bit, being correct in half of the times. So, the adversary has $3/4$ probability of answering correctly.  Assuming that the success probability at each round is independent of previous successes, the total probability of success for an adversary is $(3/4)^n$ for $n$ rounds. 
\end{proof}

In \cite{KimAKSP-2008-icisc}, an active attack against the Tu's and Piramuthu's protocol is proposed. The core idea consists of the attacker toggling a bit sent by the reader in the rapid bit exchange, while leaving the response untouched. The attacker observes the reader's response and derives a bit of the long term secret key. To prevent this attack, the message $t_B$ is incorporated by Kim et al. in their proposed scheme (Fig. 2).

\section{Kim et al.'s Distance Bounding Protocol}
Kim et al. \cite{KimAKSP-2008-icisc} proposed two authentication protocols. They proposed both a simple as well as a more efficient variant of their protocol. For ease of exposition, we shall only describe the simpler scheme. The reader should note that our proposed attack is equal effective against any of these two schemes since they are based on the same assumptions. The basic protocol  (Fig. 2) is composed of three phases:

\begin{figure*}
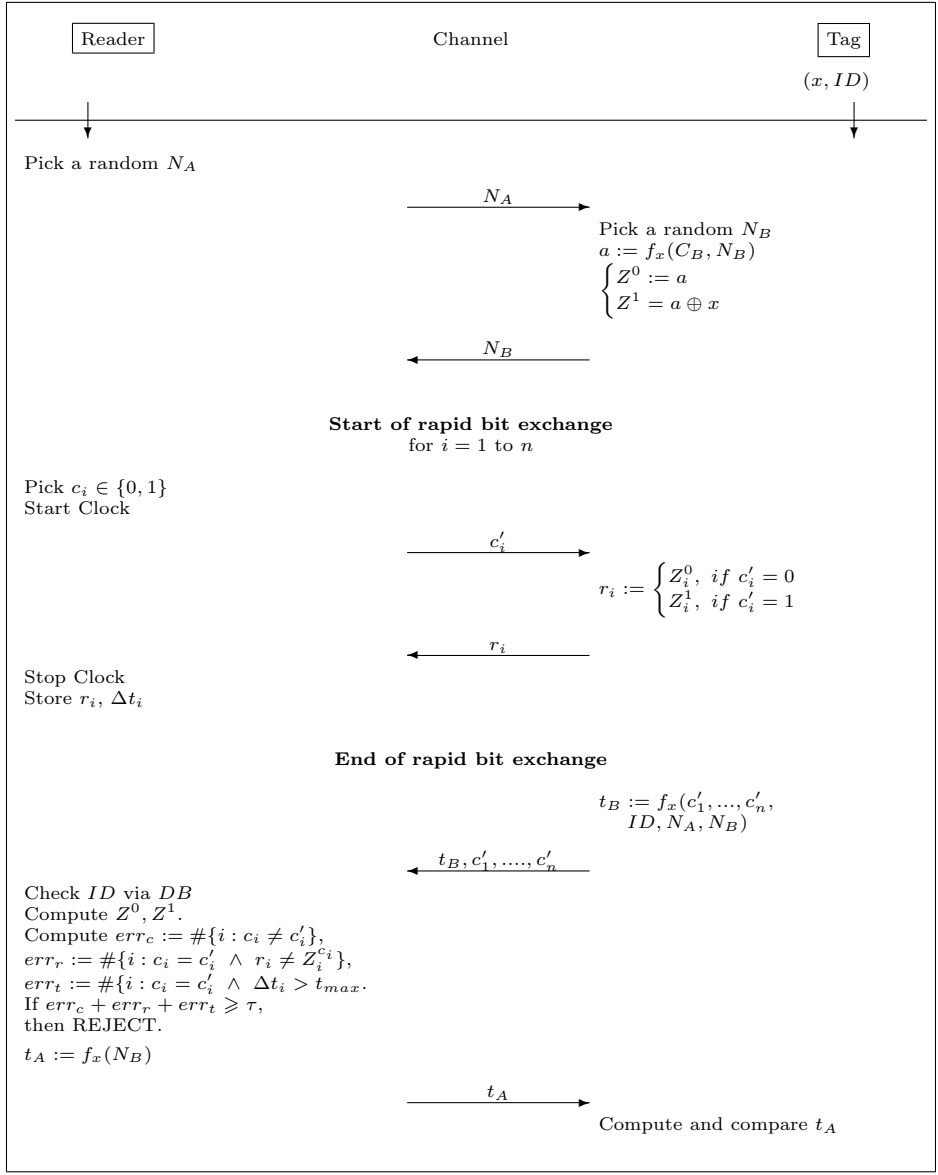

\centering
\begin{scriptsize}
\protocol{
\protheader{Reader}{\leer}{\leer}{Tag}{$(x, ID)$}
\protline{\raggedright Pick a random $N_{A}$ }{\leer}{\leer}
\protline{\leer}{\protrightarrow{$N_{A}$}}{\leer}
\protline{\leer}{\leer}{
\raggedright Pick a random $N_{B}$ \\
\raggedright $a:=f_{x}(C_B, N_{B})$ \\
\raggedright  $\begin{cases}Z^{0}:=a\\
					    Z^{1}=a \oplus x\end{cases}$}\\
\protline{\leer}{\protleftarrow{$N_{B}$}}{\leer}
\begin{center}
\textbf{Start of rapid bit exchange}\\
for $i=1$ to $n$
\end{center}
\protline{\raggedright Pick $c_{i}\in\{0,1\}$\\
	      \raggedright Start Clock}{\leer}{\leer}
\protline{\leer}{\protrightarrow{$c'_{i}$}}{\leer}
\protline{\leer}{\leer}{\raggedright $r_{i}:=
\begin{cases} Z_{i}^{0}, ~if ~ c'_{i}=0\\
Z_{i}^{1},~ if ~ c'_{i}=1\end{cases}$
}
\protline{\leer}{\protleftarrow{$r_{i}$}}{\leer}
\protline{\raggedright Stop Clock\\
	      \raggedright Store $r_i$, $\Delta t_i$}{\leer}{\leer}
\begin{center}
\textbf{End of rapid bit exchange}\\
\end{center}
\protline{\leer}{\leer}{\raggedright $t_B :=f_x(c'_1,..., c'_n,$\\
 \raggedright ~~~~$ID, N_A, N_B)$}
\protline{\leer}{\protleftarrow{$t_B, c'_1,...., c'_n$}}{\leer}
\protline{\raggedright Check $ID$ via $DB$\\
          \raggedright Compute $Z^{0}, Z^{1}$.\\
	      \raggedright Compute $err_{c}:= \#\{i:c_{i}\neq c'_{i}$\},\\
	      \raggedright $err_{r}:= \#\{i:c_{i}= c'_{i}~\land ~r_{i}\neq Z_{i}^{c_{i}}$\},\\
	      \raggedright $err_{t}:= \#\{i:c_{i}= c'_{i}~\land ~\Delta t_{i} > t_{max}$.\\
	      \raggedright If $err_{c}+err_{r}+err_{t}\geqslant \thr$,\\
	      \raggedright then REJECT. 
	      } {\leer}{\leer}	
\protline{\raggedright $t_{A}:=f_{x}(N_{B})$}{\leer}{\leer}
\protline{\leer}{\protrightarrow{$t_{A}$}}{\leer}
\protline{\leer}{\leer}{\raggedright Compute and compare $t_{A}$}}
\label{fig::figKim}
\caption{Swiss-Knife RFID Distance Bounding Protocol}
\end{scriptsize}
\end{figure*}

\begin{itemize}
\item \textbf{Preparation Phase:} The RFID reader first chooses a random number $N_A$ and transmits it to the tag. On receiving it, the tag chooses a random number $N_B$ and computes a temporary key $a:=f_x(C_B, N_B)$, where $x$ is the permanent secret key and $C_B$ is just a system-wide constant. The tag then splits its permanent secret key $x$ into two shares by computing $Z^0:=a$ and $Z^1:= a \oplus x$. Finally, the tag transmits $N_B$ to the reader.

\item\textbf{ Rapid Bit Exchange:} This phase is repeated $n$ times, with $i$ varying from $1$ to $n$, and the challenge-response delay is measured for each step.  The reader starts by choosing a random bit $c_i$, initializing the clock to zero and transmitting $c_i$ to the tag. The values received by the tag are denoted by $c'_i$.  Next, the tag answers by sending $r'_i:=Z_i^{c'_i}$. The values received by the reader are denoted by $r_i$.  On receiving $r_i$, the reader stops the clock and stores the received answer and the delay time $\Delta t_i$.

\item \textbf{Final Phase:} The tag computes $t_B := f_x(c'_1, ... , c'_n, ID, N_A, N_B)$. On receiving it, the reader performs an exhaustive search over its tag database until it finds a pair ($ID$, $x$) that matches up with the received value $t_B$. The reader computes the values $Z^0$ and $Z^1$ and checks the validity of the responses sent during the rapid bit exchange.  If errors ($err_c+err_r+err_t$) are below a threshold $\thr$, the tag is authenticated. Finally, in cases in which reader authentication is also demanded, the reader computes $t_A:=f_x(N_B)$ and sends it to the tag.
\end{itemize}

The authors argue that the security bound against mafia fraud attacks is $(\frac{1}{2})^n$. Regarding the terrorist fraud attacks, they evince that this is bounded by  $(\frac{1}{2})^v$ assuming that the adversary knows at least $n$-$v$ bits of the long-term secret key. Finally, the authors claim that privacy is guaranteed since no confidential information is transmitted on the clear. However, Kim et al. are not aware that the two versions of the protocol, as shown in Sections \ref{sec::secattack} and \ref{sec::experiments}, are vulnerable to a passive attack which compromises all the above objectives.

\section{A Full Disclosure Passive Attack} \label{sec::secattack}

In this section we present a passive attack against Kim et al.'s \cite{KimAKSP-2008-icisc} protocol which takes advantage of the weak protection mechanism adopted against terrorist fraud attacks. The described attack is not only applicable to \cite{KimAKSP-2008-icisc}  but it is also exploitable against its two predecessors \cite{TuP-2007-rfidtechnology,reid2007}.\\

\begin{assumption}
Our main assumption is that the random numbers generated by a tag do not have a long bit length (i.e. 64 or 80 bits). \\
\end{assumption}

The above assumption has two main justifications:

\begin{itemize}
  \item \textbf{Tag Resources:} Low-cost RFID tags have widespread commercial adoption \cite{Bevan2007}. Such tags have severe resource constrains and support an on-board Pseudorandom Number Generator (PRNG) with length 16 or 32 bits (e.g. Gen-2 tags \cite{EPCC1G2-2008}).
  \item \textbf{Protocol Description:} In the original paper, the authors do not specify the length of the variables used. However, we can deduce it from how the messages are built: 1) The long-term secret key $x$ and the temporary key $a$ have the same bit length due to the use of the bitwise operator in the computation of $Z^{0}$. This length is fixed to $n$ and represents the number of bits exchanged during the rapid bit exchange phase. Thus, we may conclude that the PRNG function used to generate the temporary key outputs $n$ bits at each invocation; 2) Besides the temporary key computation, the tag has to generate a random number ($N_B$ in the protocol description). For that computation the use of the PRNG function seems the most convenient option to conform with the demanding hardware restrictions of the tags.

      From 1) and 2), we may conclude that the random number $N_B$, generated by the tag has the same bit length as the number of bits ($n$ in the protocol description) transmitted during the rapid bit exchange phase.  Summarizing,  if we need to implement the protocol, we can assume that all the variables in the protocol -- long-term secret key $x$, session key $a$ and random number $N_B$ -- have a length of $n$ bits.

\end{itemize}

\begin{remark}
In Kim et al.'s \cite{KimAKSP-2008-icisc} protocol, the correct selection of the parameter $n$ -- number of bits exchanged during the rapid bit exchange phase -- determines the feasibility of the protocol. Although high values of $n$ (i.e. $n=64$ or $n=80$) are more secure, they are impractical due to the scarce resources in RFID tags \cite{AvoineT-2009-isc}. So, the $n$'s bit length is restricted to inferior values (e.g. 32 bits).
\end{remark}

Kim et al. proposed that the tag sends $Z^0_i:=a_i$ or $Z^1_i:=a_i \oplus x_i$  when receiving the challenge $r_i = 0$ or  $r_i = 1$ respectively. Knowing $Z^0$ and $Z^1$ makes it easy to calculate the value of the long-term secret key $x$, since  $Z^0 \oplus Z^1 = x$. The authors argue that the use of  $Z^0$ and $Z^1$ frustrates terrorist attacks since dishonest tags ($\overline{T}$) can not transmit both values to an intruder ($I$). Nevertheless, we show that a passive attacker can discover the long-term secret key $x$ without requiring any collaboration with $\overline{T}$ or alteration/forwarding of the messages transmitted on the channel.

\begin{theorem}
In Kim et al.'s \cite{KimAKSP-2008-icisc} protocol, a passive attacker can derive the long-term secret key of the tag (the prover) by eavesdropping on the channel over multiple executions of the authentication protocol.
\end{theorem}

\begin{proof}

Let us suppose that $n=1$, meaning that the challenges-responses transmitted during the rapid bit exchange phase have a length of 1 bit. The attacker performs the following steps:
\begin{enumerate}
  \item  He eavesdrops one authentication session. He identifies the session by the random number $N_B$ sent by the tag just before the start of the rapid bit exchange phase.  Then, he stores the bits $\{c_i, r_i\}$ transmitted (on the clear) during the rapid bit exchange phase.
  \item  He eavesdrops a new authentication session. If the value $N^*_B$ is equal to the value stored in Step 1, then the attacker stores $\{c^*_i, r^*_i\}$ and jumps to the next step. Otherwise, he repeats this step.
  \item  He checks the non equality between $c_i$ and $c^*_i$. If $c_i \neq c^*_i$, then $x_i = r_i \oplus r^*_i$ (i.e. $x_i = a_i \oplus (a_i \oplus x_i) =  (a_i \oplus x_i) \oplus a_i $). Otherwise, Step 2 is repeated.
\end{enumerate}

The generalization of the attack for an arbitrary value $n$ is straightforward:
\begin{enumerate}
  \item  The attacker eavesdrops one authentication session. He identifies the session by the random number $N_B$ sent by the tag just before the start of the rapid bit exchange phase.  Then, he stores the bits $\{c_i, r_i\}_{i=1}^n$ transmitted (on the clear) during the rapid bit exchange phase.
  \item  He eavesdrops a new authentication session. If the value $N^*_B$ is equal to the value stored in Step 1, the attacker stores $\{c^*_i, r^*_i\}_{i=1}^n$ and jumps to the next step. Otherwise, he repeats this step.
  \item  For $i=1$ to $n$, he compares the non-equality between $c_i$ and $c^*_i$. If $c_i \neq c^*_i$, then $x_i = r_i \oplus r^*_i$.  The attacker stops when all the bits of $x_i$ has been disclosed; otherwise he jumps to Step 2. 
  \end{enumerate}

\end{proof}

Therefore, the attacker is able to disclose the tag's long-term secret key by eavesdropping phase several authentication sessions on the rapid bit exchange. In the next section, we analyze the number of sessions required for a successful disclosure of the tags' long-term secret key. This depends on both the number of challenge-response bits $n$ transmitted during the rapid bit exchange phase and the channel bit error (Bit Error Rate - BER).  We should note here that the presented attack is the most harmful attack that a tag can suffer since its success enables compromises (i.e. on privacy, traceability) and further attacks (i.e. mafia and terrorist fraud attacks, etc.).

\section{Experimental Results}\label{sec::experiments}
In this section, we calculate the number of sessions that need to be eavesdropped for a successful attack.  We start by considering the simple scenario in which there are no transmission errors in the channel.  Then, we adopt a more realistic approach and consider that transmission errors can occur both in the backward (tag-to-reader) and in the forward channel (reader-to-tag).

\subsection{Ideal Communication Channel}\label{sec::idealcomm}

We could start implementing directly the attack presented in the above section. However, this attack is not very efficient since the average number of sessions required to see the same random number $N_B$ is $2^{-n}$, where $n$ is the bit length of the variable $N_B$. It is important to note that the efficiency of the attack can be increased dramatically if instead of focusing on an unique random number $N_B$ we use information from all of the eavesdropped sessions. We are actually creating a dictionary. This dictionary stores the random number $N_B$ of each session. The bits $\{c_i, r_i\}_{i=1}^n$ transmitted on the channel during the rapid bit exchange represent the meaning of each word $N_B$. The complete procedure followed by the attacker is given in detail below:

\begin{enumerate}
  \item  He initializes the meaning of the words in the dictionary to the null value (i.e. $\forall$ $N_B$, $dictionary[N_B]=null$).
  \item  He eavesdrops one session of the authentication protocol. He identifies the session by the random number $N_B$ sent by the tag:
      \begin{enumerate}
      \item He checks if the word $N_B$ exists in the dictionary. If it is not registered, he stores the bits transmitted on the channel during the rapid bit exchange (i.e. $dictionary[N_B] = \{c_i, r_i\}_{i=1}^n$). Otherwise, he jumps to the next step.
      \item He obtains the meaning of the word $N_B$ stored in the dictionary (i.e. $dictionary[NB] = \{c^*_i, r^*_i\}_{i=1}^n$).
      \item He compares the new value eavesdropped with the value stored:
      For $i=1$ to $n$, he compares the non-equality between $c_i$ and $c^*_i$. If $c_i \neq c^*_i$, then $x_i = r_i \oplus r^*_i$.
      \item He checks if all the bits of $x$ are obtained. If that is the case, he stops the loop. Otherwise, he updates the meaning of the word $N_B$ with the new eavesdropped value and jumps to Step 2. The reader should note that a list of all the values (meanings $\{c_i, r_i\}_{i=1}^n$) of each word $N_B$ could be maintained but this would increase the storage demands significatively. To increase performance, we only store the last value at the expense of loosing some effectiveness.
    \end{enumerate}
\end{enumerate}

\begin{figure}
\centering
  \includegraphics[width=8cm]{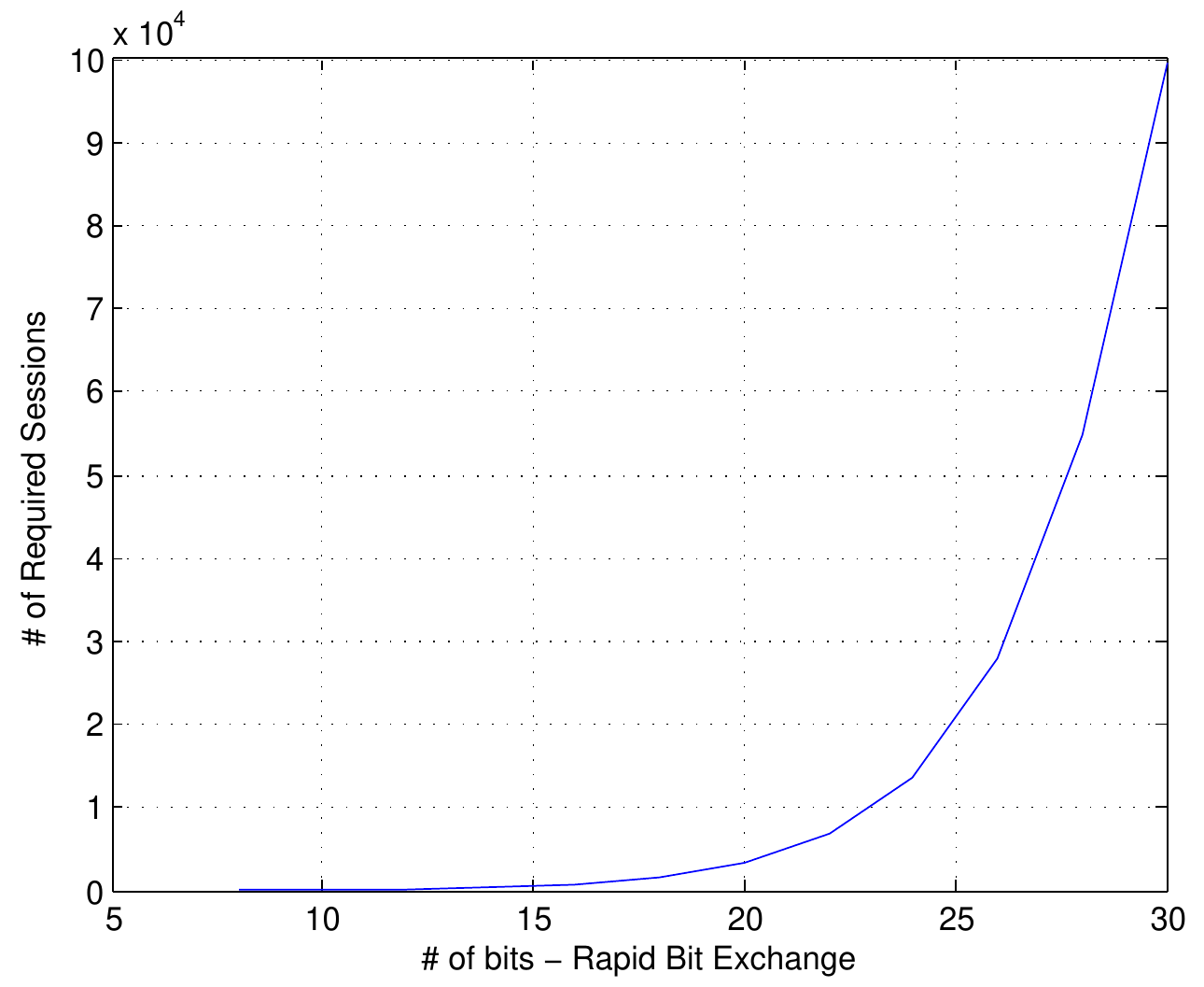}\\
  \caption{Number of Eavesdropped Sessions ($BER=0.0$)}\label{fig::fig2}
\end{figure}

We performed the above simulation for different numbers (n=\{8, 10, 12,..., 30\}) of challenge-response bits transmitted during the rapid bit exchange.  In fact, for a fixed value $n$, we repeat the simulation $2^{14}$ times to obtain an average value. The results are presented in Fig. \ref{fig::fig2}. As shown in this figure, the number of eavesdropped sessions increases when the number of bits transmitted during the rapid bit exchange increases. For example, $3,310$ and $99,526$ eavesdropped sessions are required for $n = 20$ and $n = 30$ respectively.

\subsection{Real Communication Channel}

In this section we consider that errors can appear in the channel. More precisely, we assume that errors are possible both in the backward and the forward channel. We assume that the errors are independent of each other and that the error probability is constant for all bits. Therefore,  the channel (forward or backward) produces a bit error with probability  $q$.

\begin{remark}
In this section we use the knowledge of the long-term secret key for two main purposes.  Firstly, we estimate the probability of success for an adversary that follows the same approach with that in Section \ref{sec::idealcomm} (ideal communication channel). Secondly, we reckon the average number of eavesdropped sessions required by an adversary for recovering $p*n$ bits of the secret key. The parameters $p$ and $n$ represent the percentage of bits that are recovered correctly and the bit length of the secret key respectively.

The adversary, who does not know the long-term secret key, should eavesdrop a number of session greater or equal to the number obtained in the experiments described in this section to successfully perform his attack.

\end{remark}

\begin{figure}
\centering
  \includegraphics[width=8cm]{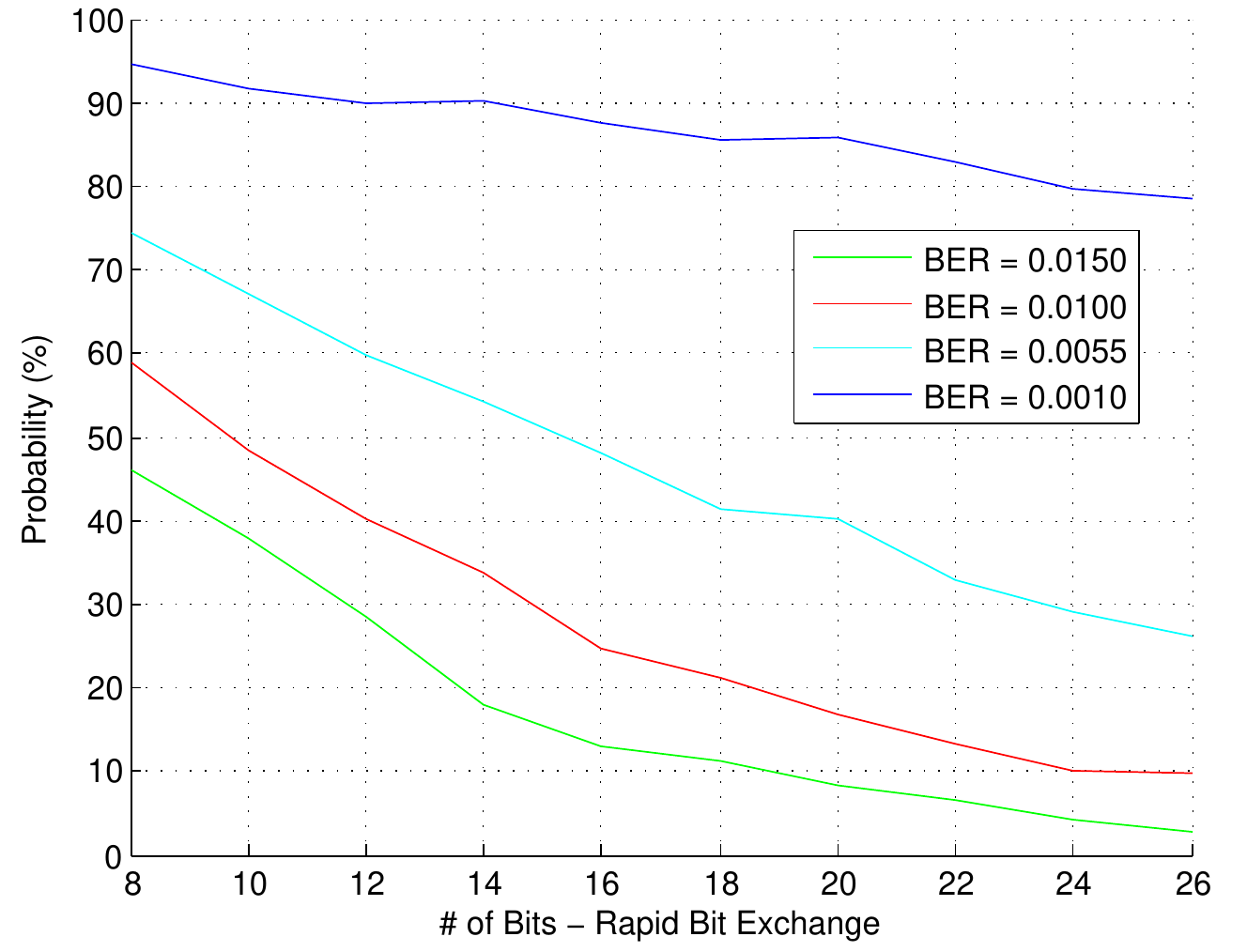}\\
  \caption{Adversary's Success Probability}\label{fig::fig3}
\end{figure}

We repeat the experiments described in the previous section but this time with the introduction of errors. Due to the errors, the probability of success is not $100\%$. That implies that some of the bits guessed may be incorrect.  Thus, if one or more bits of the key are incorrect, we consider this experiment (attack) unsuccessful and only when the whole key is revealed a success is scored.  Fig. \ref{fig::fig3} depicts the results of our simulations.  We can observe that the probability of success is over $80\%$ when the Bit Error Rate (BER) is $10^{-3}$.  On the contrary, when the BER is extremely high (BER=0.015), the probability of success declines exponentially as we increase the number of bits transmitted during the rapid bit exchange.

However, the above attack can be made twice as effective, by increasing the number of sessions that need to be eavesdropped.  For a particular probability of success $p$, we can estimate the average number of required eavesdropped sessions by means of the experiment explained below. In fact, the probability $p$ can be interpreted as the percentage of bits that coincided between the searched key and the estimated key. The complete procedure followed by the attacker is given in detail below:

\begin{enumerate}
  \item  He initializes the meaning of the words in the dictionary (i.e. $\forall$ $N_B$, $dictionary[N_B]=null$), as well as, the list of possible keys to the null value. 
  \item  He eavesdrops one session of the authentication protocol. Each session is identified by the random number $N_B$ sent by the tag.
      \begin{enumerate}
      \item The attacker checks if the word exists in the dictionary. If it is not registered, he stores the bits transmitted on the channel during the rapid bit exchange (i.e. $dictionary[NB] = \{c_i, r_i\}_{i=1}^n$). Otherwise, he jumps to the next step.
      \item He obtains the meaning of the word $N_B$ stored in the dictionary (i.e. $dictionary[NB] = \{c^*_i, r^*_i\}_{i=1}^n$).
      \item He compares the new eavesdropped value with the value stored: For $i=1$ to $n$, he compares the non-equality between $c_i$ and $c^*_i$. If $c_i \neq c^*_i$, then $x_i = r_i \oplus r^*_i$.
      \item He checks if all the bits of $x$ are already disclosed. If they are, he stops the loop and jumps to Step 3. Otherwise, he updates the new meaning of the word with the new  eavesdropped value and jumps to Step 2.
             \end{enumerate}
   \item Key searching:
   \begin{enumerate}
     \item The attacker stores the new key in the list of possible keys.
     \item He derives the most common value that appears in the list of possible keys. 
     \item He compares the most common value of the key (derived from the previous step) with the pursued key (see Remark 1). If the differences (in bits) between these two values is lower than $p * n$, the whole process stops. Otherwise, he jumps to Step 2 again.
    \end{enumerate}
\end{enumerate}

We perform a careful study of the number of sessions required, by following the above experimental procedure.
As previously,  we perform  $2^{14}$ independent trials to obtain an average value. Fig. 4-6 summarize the results obtained.

\begin{figure}
\centering
  \includegraphics[width=8cm]{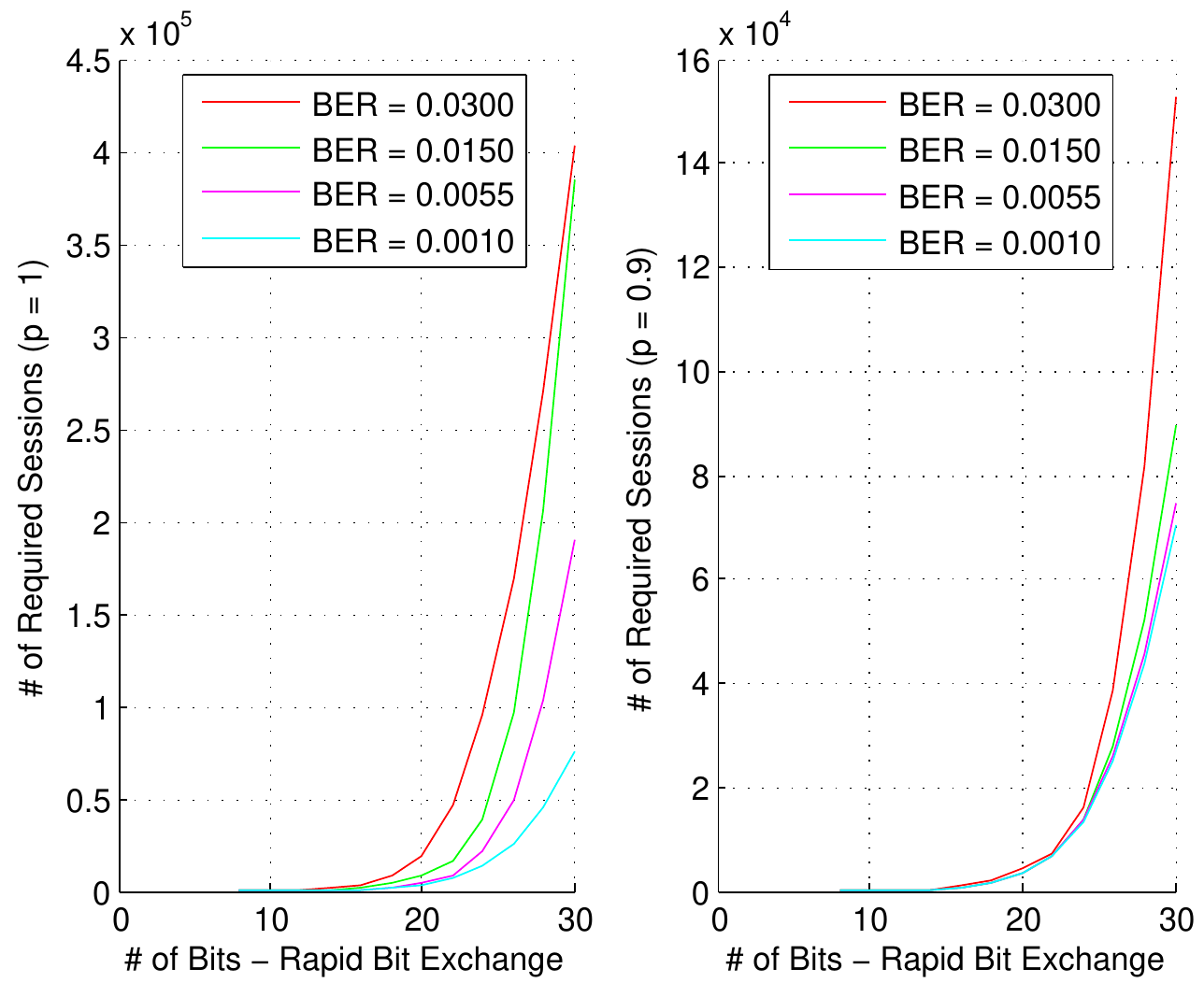}\\
  \caption{Number of Eavesdropped Sessions (p = 1 and p = 0.9)}\label{fig::fig4}
\end{figure}

Fig. \ref{fig::fig4} depicts the number of sessions required for $p=1$ and $p=0.9$. By comparing these results, we observe that the number of sessions is increased by an order of magnitude when the percentage of recovered bits of the key is increased from $90\%$ to $100\%$.  For $p=0.9$, the influence of errors in the channel is only slightly noticeable when the BER is extremely high. However, for $p=1$, the effect of the BER is marked. As expected, the number of required sessions increases when the number of challenge-response bits transmitted during the rapid bit exchange raises and/or the number of errors in the channel (BER) increases.

\begin{figure}
\centering
  \includegraphics[width=8cm]{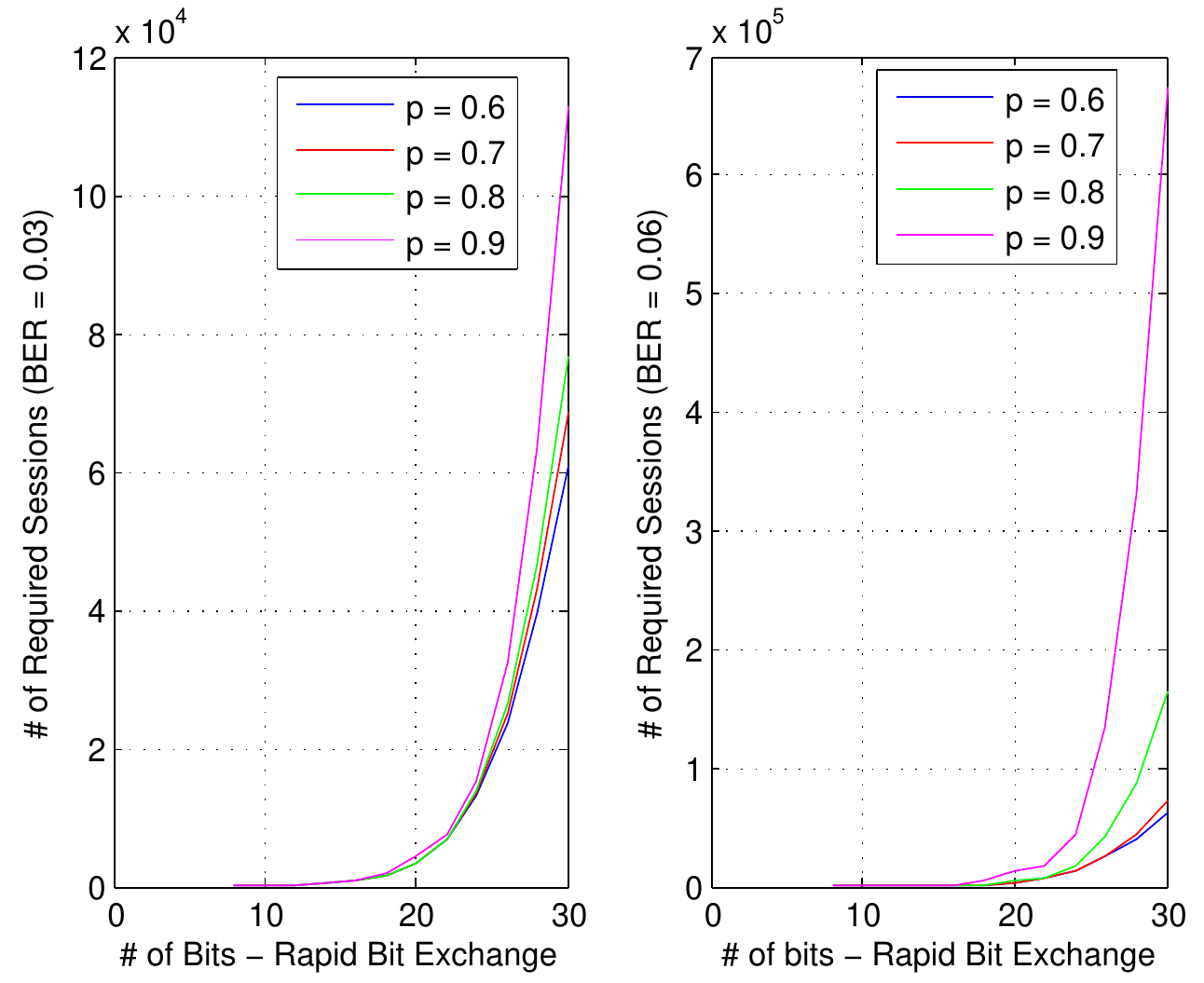}\\
  \caption{Number of Eavesdropped Sessions (BER = 0.03 and BER = 0.06)}\label{fig::fig5}
\end{figure}

Fig. \ref{fig::fig5} depicts the number of required eavesdropped sessions for very noisy channels. Specifically, we show how the number of required eavesdropped sessions changes as the number of bits increases for probability $p$ ($p=\{0.6, 0.7, 0.8, 0.9\}$) and BER values (BER = 0.03 or BER = 0.06).  For a moderately high BER = 0.03, there is no significant difference between the number of sessions required for recovering a part of the key (i.e. 60-70$\%$) or almost the whole key (i.e. 90$\%$). On the contrary, this difference is evident for higher  BER (i.e. 0.06 in the Fig. \ref{fig::fig5}). From these figures we conclude that when the BER is low, the additional effort required to obtain a larger percentage of the key is small.

\begin{figure}
\centering
  \includegraphics[width=8cm]{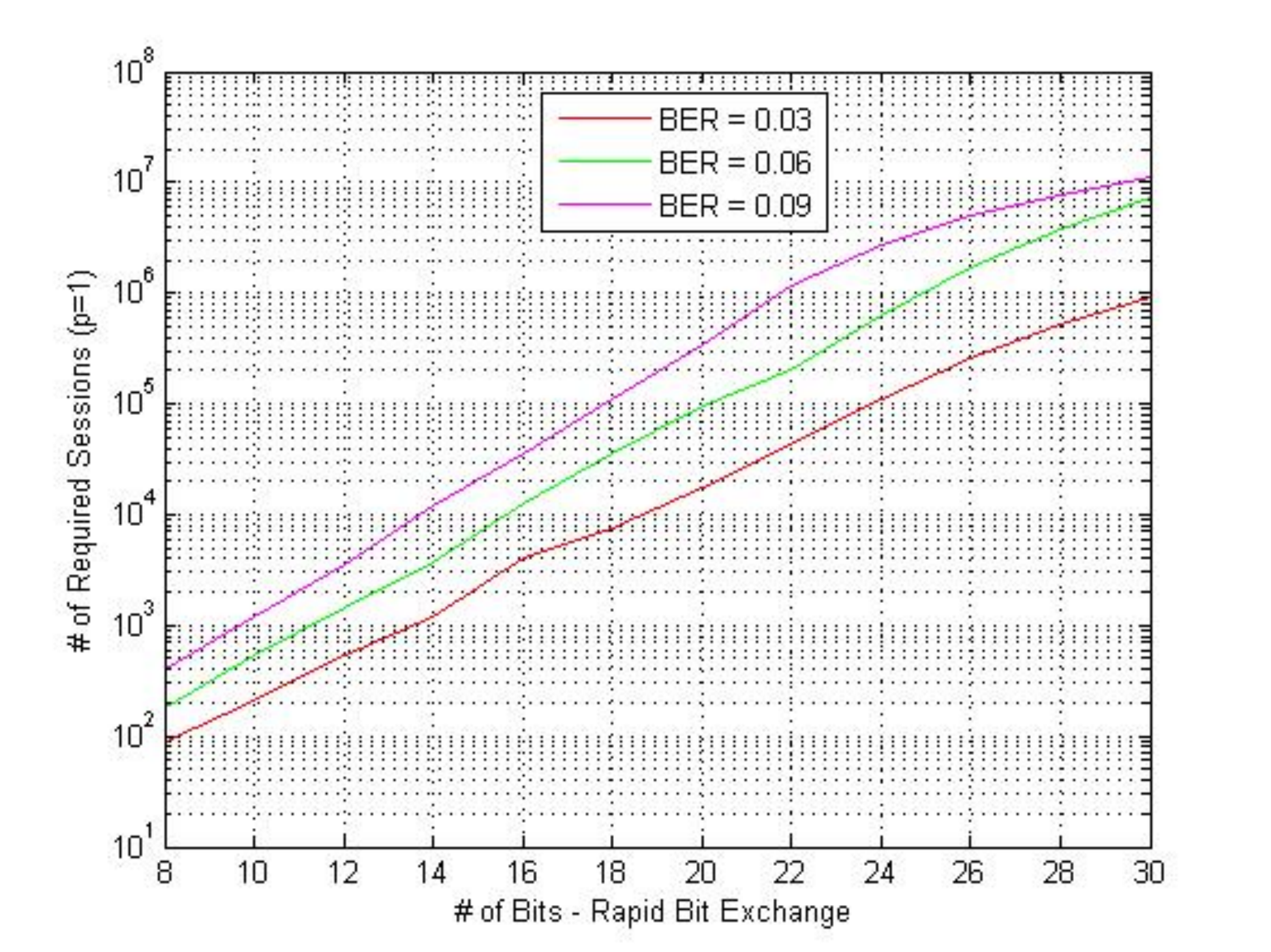}\\
  \caption{Number of Eavesdropped Sessions (BER = \{0.03, 0.06, 0.09\})}\label{fig::fig6}
\end{figure}

The effect of noise becomes clearer in Fig. \ref{fig::fig6}, which shows the number of required eavesdropped sessions to fully obtain the key ($p = 1$), versus the number of bits in the bit exchange phase, for three different BER values (BER = $\{0.03, 0.06, 0.09\}$).  As a rule of thumb, we observe that when the BER is twice as much as was previously simulated, the number of required session is multiplied at least by a factor of three.  Nevertheless, even for an extremely high BER = 0.03, the number of required eavesdropped sessions is inferior to the number of attempts that an adversary would need to perform a brute force attack\footnote{If the number of bits transmitted during the rapid bit exchange phase is $n$, then the attacker has to guess a word of $n$ bits.} (i.e. $\frac{\sharp~of~Required~Sessions}{2^n} < 1$).\\

\begin{remark}
As we already mentioned before, the feasibility of the described attack is significantly superior to that of a brute force attack. As a rule of thumb, $2^{\frac{n}{2}}<<2^n$, where $n$ represents the number of challenge-response bits transmitted during the rapid bit exchange phase. On the other hand,  we are aware of the main drawback of this passive attack.  Specifically,  the attack requires the tag to be engaged in many sessions; something that is not required for the brute force attack.
\end{remark}

\begin{remark}
In \cite{AvoineBKLM-2009-eprint} a formal framework to cryptanalyse distance bounding protocols is presented. According to this proposal, our attack can be classified as a ``no-ask'' strategy since the adversary does not interact at all with the prover during the attack.
\end{remark}

\section{Design Guidelines}\label{sec::guidelines}

In this section we describe the standard procedures that protocol designers should consider to propose a secure and efficient RFID protocol against relay attacks. This is the first time -- to the best of our knowledge -- that a complete set of guidelines is presented. Our work complements the formal framework for cryptanalyzing distance-bounding protocols proposed by Avoine et al. \cite{AvoineBKLM-2009-eprint}.  Particularly, we focus on distance-bounding protocols in which the RFID reader (prover) sends a single-bit challenge and the tag (verifier) replies with a rapid single-bit response. The above procedure is repeated $n$ times, where $n$ represents a security parameter.  Finally, the reader computes an upper-bound of the distance between both entities by measuring the delay time between the challenges and responses, which is based on the fact that messages cannot propagate faster than light. We emphasize that the tag should send its answer immediately after receiving a challenge from the reader for the bound to be tight.

\subsection{Distance Fraud Attacks}\label{sec::dfsolution}
 A distance fraud attack is possible when there is no relationship between the challenge bits and the response bits exchanged during the distance verification.  If a fraudulent tag $\overline{T}$ knows when the challenge bits  will be sent,  it can trick the reader $R$ by sending the response bits before receiving the challenge bits.  Thus, $R$ computes a wrong upper bound regarding its physical distance to $\overline{T}$. To resolve this problem, we present three solutions. Of those, we mostly recommend the third as it is the most feasible.

\begin{itemize}
\item \textbf{Solution A:} The RFID reader sends challenge bits at random chosen times.  Due to this countermeasure, $\overline{T}$ cannot send response bits before he has received the challenge bits as $\overline{T}$ can not predict when the reader will expect a response. Brands and Chaum \cite{brands94} suggested that  it is sufficient for $R$  to send its response randomly at one of two discrete times (i.e. each $3i$ or $3i+1$ clock cycles). This strategy has a success probability of $(\frac{1}{2})^n$ if the selection of discrete times is random.

\item \textbf{Solution B:} The RFID reader can use void challenges \cite{munilla2} to detect that $T$ is not waiting to receive the challenge bits. A void challenge is a challenge which the reader intentionally leaves without sending. That is, the challenge bits  $c_i$ sent by the reader can take three different values $\{0, 1, void  \}$. If the reader detects that a response bit is received during the interval of a void challenge, the dishonest tag $\overline{T}$ is detected.  As previously, the strategy has a success probability of $(\frac{1}{2})^n$. Note that the inclusion of void challenges is equivalent to transmitting the challenge bits $c_i=\{0, 1\}$ at randomly chosen times.
\item \textbf{Solution C:} The tag $T$ must select its response depending on the challenge sent by the reader $R$.  A possible scheme is presented below:
    \begin{itemize}
      \item \textbf{Step 1:} $R$ generates at random $n$ bits $c_i$.
      \item \textbf{Step 2:} $T$ generates at random $n$ bits $m_i$ and commits these bits using a secure commitment scheme (i.e. $commit\{m_i,...,m_n\}$).\footnote{As an alternative solution, Brands and Chaum \cite{brands94} suggested to create a public bit string $\{m_i,...,m_n\}$, where the choice of bits $m_i$ is irrelevant.}
       \item \textbf{Step 3:} The rapid bit exchange phase can start. This phase is repeated $n$ times, with $i$ varying from 1 to $n$, and the challenge-response delay is measured for each step:
           \begin{itemize}
             \item $R$ sends the bit $c_i$ to $T$ and initializes the clock to zero.
             \item $T$ sends the bit $r_i = c_i \oplus m_i$ immediately after he receives $c_i$.
             \item On receiving $r_i$, $R$ stops the clock and stores the delay time.
           \end{itemize}
      \item \textbf{Step 4:} $T$ opens the commitment and $R$ verifies whether $\{ {c_i \oplus r_i}\} _{i=1}^{n}$ equals $\{{m_i}\}_{i=1}^n$. If so, $R$ computes an upper bound on the distance to $T$ using the maximum of the measured delay times.
    \end{itemize}

     Similarly to solution A the success probability for an adversary is at most $(\frac{1}{2})^n$ \cite{brands94}.
\end{itemize}

\begin{theorem}\label{th::distance}
If an RFID tag is not in close proximity to an RFID reader then, independently of the selected strategy $\{A, B, C\}$ to avoid distance frauds, the success probability for an adversary to launch a distance fraud attack is at most $(\frac{1}{2})^n$ \cite{brands94}.
\end{theorem}

\subsection{Mafia Fraud Attacks}

Hancke and Kuhn \cite{hancke05}  proposed a distance bounding protocol which incorporates the use of a rapid bit exchange phase. In this protocol, the success probability with which a mafia attack can succeed is bounded by $(\frac{3}{4})^n$ \cite{TuP-2007-rfidtechnology}.  However, this probability is superior than the optimal value $(\frac{1}{2})^n$.   To achieve this optimal value, as suggested in \cite{KimAKSP-2008-icisc,brands94}, the tag has to sign or encrypt the bits sent back and forth during the rapid bit exchange. A possible scheme is presented below, where $R$ and $T$ denote the reader and the tag respectively.
    \begin{itemize}
      \item \textbf{Step 1:} $R$ generates at random $n$ bits $c_i$.
      \item \textbf{Step 2:} $T$ generates at random $n$ bits $r_i$.
       \item \textbf{Step 3:} The rapid bit exchange phase can start. This phase is repeated $n$ times, with $i$ varying from 1 to $n$, and the challenge-response delay is measured for each step:
           \begin{itemize}
             \item $R$  sends $c_i$ to $T$ and initializes the clock to zero.
             \item $T$ sends the bit $r_i = c_i \oplus m_i$  to the reader $R$ immediately after he receives $c_i$.
             \item On receiving $r_i$, $R$ stops the clock and stores the received value and the delay time.
           \end{itemize}
      \item \textbf{Step 4:} $T$ concatenates the challenges and responses to create a message $\{c_1 || c_2 || ... || c_n || r_1 || r_2 || ... || r_n \}$ of length $2n$. Then, he signs or encrypts the resulting message, and sends the result to $R$.
    \end{itemize}

The reader $R$ determines an upper bound on the distance to the tag using the maximum of the delay times measured.  Finally, he accepts this value once the correctness of the challenge-response bits are checked by using the last encrypted/signed message.

\begin{theorem}\label{th::mafiafraud}
If the signature/encryption scheme is secure and the tag is not in physical proximity to the reader, an adversary has a success probability upper bounded by $(\frac{1}{2})^n$ \cite{KimAKSP-2008-icisc,brands94}.
\end{theorem}

The main drawback of using a final signature at the end of the protocol is the fact that this message has to be sent by normal communication method with error detection or correction technique \cite{KimA-2009-cans}. The execution time of the whole protocol is thus increased in comparison with protocols that omit this last message.   However, the existing alternatives (void-challenges \cite{munilla2}, predefined challenges \cite{KimA-2009-cans}, tree-based approach \cite{AvoineT-2009-isc} and multistate enhancement \cite{AvoineFM-2009-indocrypt}) have other relevant drawbacks such as excessive memory requirements and performance problems (i.e. complex coding schemes and greater bandwidth requirements).

\subsection{Terrorist Fraud Attacks}
Bussard \cite{Bussard-2004-thesis,BussardB05} suggested a mechanism through which it is impossible to mount a successful terrorist fraud attack unless the attacker discloses the private key in some way.  Although Bussard's scheme is correct, it can not be applied to devices with limited resources since it relies on public key cryptography.  Some alternatives have been proposed in the context of low-cost RFID tags but all of them are vulnerable to a full disclosure attack conducted by a passive attacker, as shown in Sections \ref{sec::secattack} and \ref{sec::experiments}. As a solution, we present two schemes that offer protection against terrorist attacks and conform to the requirements of severely resource constrained devices such as RFID tags.  Case A assumes that tags support an encryption function ($E$) \cite{BogdanovKLPPRSV-2007-ches,HellJM-2006} and a Pseudo-Random Function - PRF ($f$) \cite{lee-prf-2007}. In case B, tags only have a PRF  and can compute simple bitwise operations. These two schemes are described below; where the long-term secret key of the tag is denoted by  $x$:
     \begin{enumerate}
      \item \textbf{Preparation Phase:}  The tag computes a temporary key (e.g. $ a = f_x(N_{R}, N_{T}, W)$), where $N_{R/T}$ denotes a random number generated by the reader ($R$) / tag ($T$) and $W$ represents any additional parameter.  The tag then splits its permanent secret key $x$ into two shares by computing:
          \begin{itemize}
            \item[--]Case A: $Z^0:=a$ and $Z^1:= E_a(x)$.
            \item[--]Case B: $Z^0:=a$ and $Z^1:= f_a(N'_T, W') \oplus x$. \\
             $N'_T$ denotes a random number generated by $T$ and $W'$ any extra parameter respectively.
          \end{itemize}
      \item \textbf{Rapid Bit Exchange:} This phase is repeated $n$ times, with $i$ varying from 1 to $n$.  At every step $i$, we measure the challenge-response delay time.
    \begin{itemize}
             \item \textbf{Step 1:} $R$  generates a random bit $c_i$, initializes the clock to zero and transmits $c_i$ to $T$.
             \item \textbf{Step 2:} $T$ sends the bit $r_i:=Z_i^{c_i}$ to the reader $R$ immediately after he receives $c_i$.
             \item \textbf{Step 3:} On receiving $r_i$, $R$ stops the clock and stores the delay time.
    \end{itemize}
    \end{enumerate}

\begin{theorem}\label{th::terrorist}
If a secure encryption function and/or a pseudo-random function is used, the probability with which a terrorist fraud attack can succeed is bounded by $(\frac{3}{4})^v$, when $n-v$ bits of the long-term secret key are revealed from the dishonest tag $T$ to the terrorist tag $\overline{T}$.
\end{theorem}

\begin{proof}
Let us suppose that the attacker knows  $n-v$ bits of the secret key. Additionally, we can assume the worst case scenario in which the dishonest tag $T$ transmits the whole $n$ bits of $Z^0$ or $Z^1$ to $\overline{T}$. In this situation and during the rapid bit exchange, $T$ may replay incorrectly  for any of the $v$ (from a total of $n$)  bits for which it has not obtained the corresponding secret key bits. For those cases,  in half of the times he receives $r_i=0$ and therefore $T$ knows the response in advance as $Z^0$ was completely revealed.  In the other half of the cases ($r_i=1$), $T$ can send a guessed bit, being correct half of the times (in the worst case). Thus, the adversary has $3/4$ probability of answering correctly. Finally, assuming that the success probability at each round is independent of previous successes, the total probability of success is upper bounded by $(\frac{3}{4})^v$, in case $n-v$ bits of the secret key are revealed. 
\end{proof}

\subsection{Dictionary Attack}
As a consequence of splitting the key to offer protection against terrorist attacks, a dictionary attack may be conducted by a passive attacker as presented in Sections \ref{sec::secattack} and \ref{sec::experiments}.  In this attack, the adversary takes advantage of observing multiple sessions in which the same temporary key -- random numbers associated to the current session -- is used.  Specifically, the feasibility of the proposed attack is due to the bit length of the random numbers generated by the tags.  In standard cryptography, bit lengths of 64 or 80 bits are used. However, in limited devices (i.e. low-cost RFID tags, sensor nodes, etc.), the bit length of variables is drastically reduced due to resource constraints. Thus, the usage of several temporary keys and random numbers of small bit length -- associated to each session -- is necessary to thwart this kind of attacks.

\section{The Hitomi RFID Distance Bounding Protocol}

In this section, a new RFID distance bounding protocol is presented. The proposed protocol follows the guidelines described in the previous section and attempts to offer resistance to the most common attacks (i.e. distance fraud, mafia fraud, terrorist fraud and dictionary attacks).  Our proposal is not {\em ex nihilo}, but inspired by the Swiss-Knife RFID distance bounding protocol \cite{KimAKSP-2008-icisc}. The organization of this section is as follows.  First, a description and analysis of the protocol is presented. Then, we perform a study of the threshold/errors defined in the final phase of the protocol. These errors are similar to those defined in the original scheme but a rigorous analysis of these values is missing from the original proposal.

\subsection{The Protocol}
\begin{figure*}
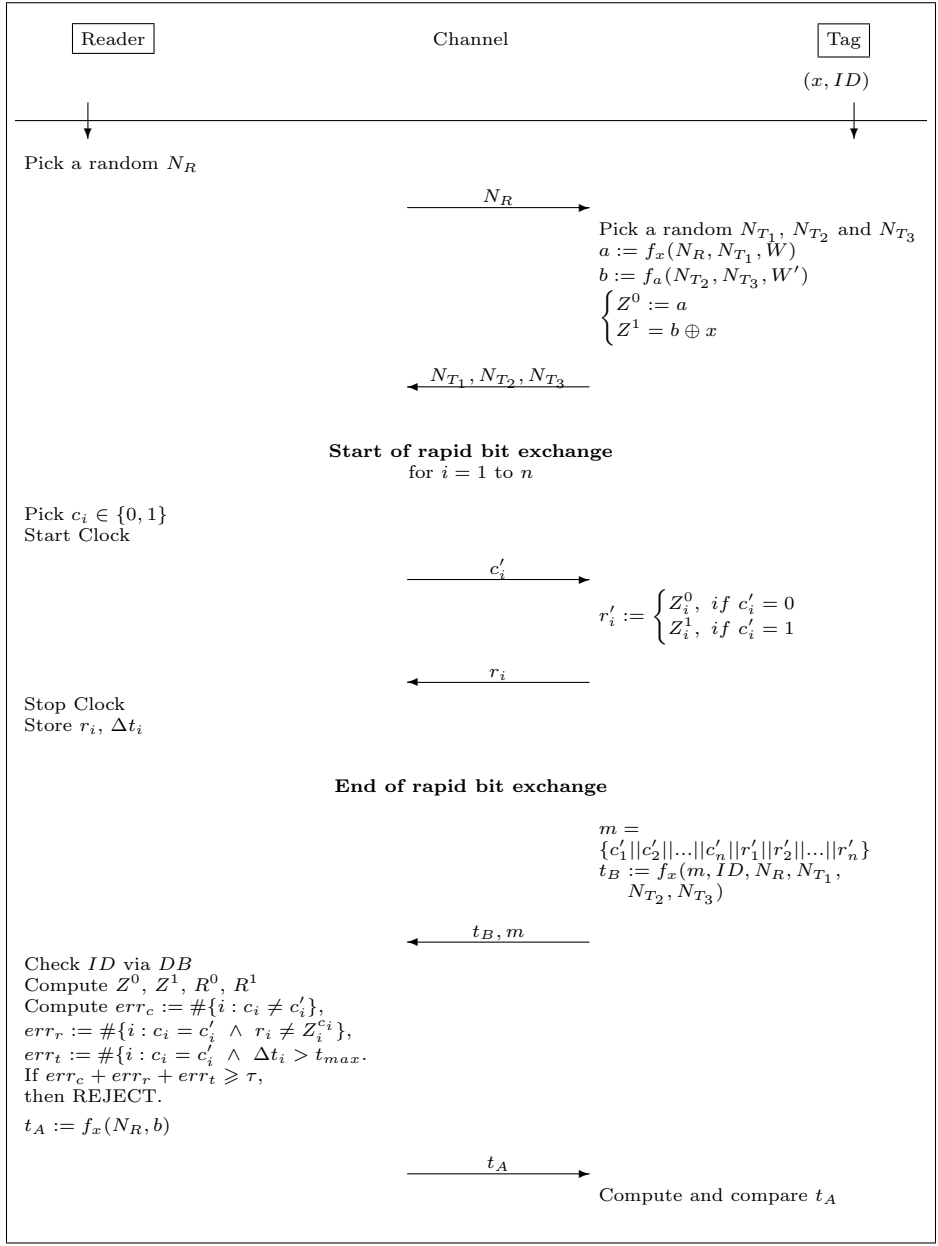

\centering
\begin{scriptsize}
\protocol{
\protheader{Reader}{\leer}{\leer}{Tag}{$(x, ID)$}
%
%
%
%
\protline{\raggedright Pick a random $N_{R}$ }{\leer}{\leer}
\protline{\leer}{\protrightarrow{$N_{R}$}}{\leer}
\protline{\leer}{\leer}{
\raggedright Pick a random $N_{T_{1}}$, $N_{T_{2}}$ and $N_{T_{3}}$\\
\raggedright $a:=f_{x}(N_{R}, N_{T_1}, W)$ \\
\raggedright $b:=f_a(N_{T_{2}}, N_{T_{3}}, W')$ \\
\raggedright  $\begin{cases}Z^{0}:=a\\
					    Z^{1}= b \oplus x\end{cases}$}\\
\protline{\leer}{\protleftarrow{$N_{T_{1}}, N_{T_{2}}, N_{T_{3}}$}}{\leer}
\begin{center}
\textbf{Start of rapid bit exchange}\\
for $i=1$ to $n$
\end{center}
\protline{\raggedright Pick $c_{i}\in\{0,1\}$\\
	      \raggedright Start Clock}{\leer}{\leer}
\protline{\leer}{\protrightarrow{$c'_{i}$}}{\leer}
\protline{\leer}{\leer}{\raggedright $r'_{i}:=
\begin{cases} Z_{i}^{0}, ~if ~ c'_{i}=0\\
Z_{i}^{1},~ if ~ c'_{i}=1\end{cases}$
}
\protline{\leer}{\protleftarrow{$r_{i}$}}{\leer}
\protline{\raggedright Stop Clock\\
	      \raggedright Store $r_i$, $\Delta t_i$}{\leer}{\leer}
\begin{center}
\textbf{End of rapid bit exchange}\\
\end{center}
\protline{\leer}{\leer}{\raggedright $m= \{c'_1 || c'_2 || ... || c'_n || r'_1 || r'_2 || ... || r'_n \}$\\
\raggedright $t_B :=f_x(m, ID, N_{R}, N_{T_{1}},       $\\
 \raggedright ~~~~$N_{T_{2}}, N_{T_{3}})$}
\protline{\leer}{\protleftarrow{$t_B, m$}}{\leer}
\protline{\raggedright Check $ID$ via $DB$\\
          \raggedright Compute $Z^{0}$, $Z^{1}$, $R^{0}$, $R^{1}$\\
	      \raggedright Compute $err_{c}:= \#\{i:c_{i}\neq c'_{i}$\},\\
	      \raggedright $err_{r}:= \#\{i:c_{i}= c'_{i}~\land ~r_{i}\neq Z_{i}^{c_{i}}$\},\\
	      \raggedright $err_{t}:= \#\{i:c_{i}= c'_{i}~\land ~\Delta t_{i} > t_{max}$.\\
	      \raggedright If $err_{c}+err_{r}+err_{t}\geqslant \thr $,\\
	      \raggedright then REJECT. 
	      } {\leer}{\leer}	
\protline{\raggedright $t_{A}:=f_{x}(N_{R}, b)$}{\leer}{\leer}
\protline{\leer}{\protrightarrow{$t_{A}$}}{\leer}
\protline{\leer}{\leer}{\raggedright Compute and compare $t_{A}$}}
\begin{quote}
    ~~~~~~~~~~~~~~~~~~~~~~~~~~~~~\textbf{Fig. 8.}  Hitomi RFID Distance Bounding Protocol
\end{quote}

\caption{Hitomi RFID Distance Bounding Protocol} \label{fig::hitomi}
\end{scriptsize}
\end{figure*}

The tag and the reader share a long-term secret key $x$ and each tag has a unique identifier $ID$. The tags' capabilities support a Pseudo-Random Function - PRF ($f$)
and can perform bitwise operations. To avoid ambiguity, we assume that all the variables have the same bit length which is fixed by the number of challenge-response bits transmitted during the rapid bit exchange phase and denoted by $n$.  The messages exchanged in the different phases (i.e. preparation phase, rapid bit exchange phase and final phase) of the protocol (Fig. \ref{fig::hitomi}) are described below.

\begin{itemize}
\item \textbf{Preparation Phase:} The protocol starts with a preparation phase that involves invocations of the PRF, random number generations and the computation of several bitwise operations.
\begin{itemize}
  \item [--] The reader chooses a random nonce $N_R$ and transmits it to the tag.
  \item [--] The tag chooses three random numbers $\{N_{T_{1}}, N_{T_{2}}, N_{T_{3}} \}$. Then, it computes the temporary keys $\{a, b\}$ as described below:
      \begin{equation*}
      \begin{cases}
      a:=f_{x}(N_{R}, N_{T_1}, W)\\
      b:=f_a(N_{T_{2}}, N_{T_{3}}, W') \\
      \end{cases}\\
      \end{equation*}
      where $W$ and $W'$ represent any extra parameters.\\
      Then, the tag splits its permanent secret key $x$ into two shares by computing:
      \begin{equation*}
      \begin{cases}Z^{0}:=a\\
				   Z^{1}:= b \oplus x \end{cases}
      \end{equation*}
      Finally, the tag transmits $\{N_{T_{1}}, N_{T_{2}}, N_{T_{3}} \}$ and $\{W, W'\}$ to the reader.
\end{itemize}

 \item \textbf{Rapid Bit Exchange Phase:} This phase is repeated for $n$ rounds. At the $i$-th round, we measure the challenge-response delay time.
    \begin{itemize}
             \item[--] $R$  generates a random bit $c_i$, initializes the clock to zero and sends $c_i$ to $T$. We denote by $c'_i$ the value received by the tag, which may be non-equal to $c_i$ due to errors or alterations in the channel.
             \item[--] $T$ sends the bit $r'_i:=Z_i^{c'_i}$ to the reader immediately after receiving $c'_i$. Similarly, we denote by $r_i$ the value received by the reader.
             \item[--] On receiving $r_i$, $R$ stops the clock and stores the delay time.
    \end{itemize}
\item \textbf{Final Phase:} The tag concatenates the challenge and response bits to obtain $m= \{c'_1 || c'_2 || ... || c'_n || r'_1 || r'_2 || ... || r'_n \}$, where $m$'s length is equal to $2n$ bits.  Then, it computes $t_B$ by ciphering the concatenation of $m$, the tag identifier $ID$ and the random numbers involved in the preparation phase.
    \begin{equation*}
    t_B :=f_x(m, ID, N_{R}, N_{T_{1}}, N_{T_{2}}, N_{T_{3}})
    \end{equation*}
    Finally, the tag sends the pair $\{t_B, m\}$ to the reader.
    The reader checks the correctness of the values received:
    \begin{itemize}
      \item [--] The reader performs an exhaustive search in its database until a match between a pair $\{ID, x\}$ and $t_B$ is found.
      \item [--] The reader computes a local version of the temporary keys $\{a, b\}$. 
      \item [--] The reader checks the validity of the responses received during the rapid bit exchange. Specifically:
      \begin{itemize}
        \item $err_c$: it counts the number of times that $c_i \neq c'_i$.
        \item $err_r$: it counts the number of times that $c_{i} = c'_{i}$ but $r_{i}\neq Z_{i}^{c_{i}} \oplus R_{i}^{c_{i}}$.
        \item $err_t$: it counts the number of times that $c_{i} = c'_{i}$ but the response delay $\Delta t_{i}$ is above a defined time threshold $t_{max}$.
        \item Finally, it checks if $err_c + err_r + err_t $ is below a fault tolerance threshold $\thr$. If not, the protocol is aborted.
      \end{itemize}
    \item[--]In those cases in which reader authentication is demanded, a final message is exchanged. The reader computes $t_{A}:=f_{x}(N_{R}, b)$ and transmits it to the tag. Once the tag checks its correctness, the two entities are mutually authenticated.
    \end{itemize}
\end{itemize}

The protocol provides the mutual authentication between the tag (prover) and the reader (verifier). In fact, the Swiss-Knife and consequently the Hitomi protocol, both inherit the security properties of the  $MAP1$ scheme \cite{188164,1062066} on which these proposals are based.  For a detail description of those security properties the reader is urged  to consult \cite{188164,1062066}.

In order to guarantee privacy protection, we avoid transmitting identifiers in plain-text. More precisely, the tag transmits its identifier after the completion of the rapid bit exchange phase. This identifier is anonymized since it is incorporated in the computation of $t_B$ which is  based on the usage of the PRF and the whole collection of nonces linked to the current session.  A traceability attack using previous protocol executions is not possible since the values transmitted on the clear are either random numbers or the output of the PRF, which includes the above mentioned nonces as inputs. A disadvantage of this solution is that the reader has to conduct an exhaustive search in the back-end database to retrieve the identity of the tag. However, there is not known alterative method that offers this level of security \cite{KimAKSP-2008-icisc}.

The Hitomi protocol, as defined,  complies with the guidelines presented in Section \ref{sec::guidelines}. We adopt Solution C (Section \ref{sec::dfsolution}) to thwart attacker plans according to distance fraud and attempting to optimize the performance of the proposed scheme. Although Solutions A and B are as effective in thwarting distance fraud attacks as Solution C,  their implementation is more complicated. Furthermore, we employ a final signature -- a signed message $t_B$ containing the received and sent challenges -- to hinder mafia fraud attacks. We have not examined other approaches such as void-challenges and predefined challenges, since we focus on protocol simplicity and communication efficiency.

\begin{corollary}
 Based on Theorems \ref{th::distance} and \ref{th::mafiafraud}, the success probability of a mafia and distance fraud attack against the Hitomi RFID distance bounding protocol is upper bounded by  $(\frac{1}{2})^n$.
\end{corollary}

The secret key $x$ is split into two parts $\{Z^0, Z^1\}$ to combat terrorist attacks.  Basically, the above construction represents a secret sharing strategy. Each time the tag is interrogated ($c'_i$) in the rapid bit exchange phase, the tag discloses only one part of each bit ($Z_i^{c'_i}$). Thus, no information on the secret key $x$ is revealed through the responses bits $r_i$.  Additionally, there is a non-linear relation between ($Z^0$) and ($Z^1$) on the contrary with what happens in previous proposals \cite{KimAKSP-2008-icisc,TuP-2007-rfidtechnology,reid2007} in which  $x_i = Z^0_i \oplus Z^1_i$.

\begin{corollary}
Due to Theorem \ref{th::terrorist} and assuming that a secure pseudo-random function $f$ is used, the success probability of a terrorist fraud attack against the Hitomi RFID distance bounding protocol is upper bounded by $(\frac{3}{4})^v$, when $n-v$ bits of the long-term secret key are revealed from the dishonest tag $T$ to the terrorist tag  $\overline{T}$.
\end{corollary}

\begin{remark}
If we follow the strategy introduced in Section \ref{sec::secattack} against the Hitomi protocol, the dictionary attack has a complexity significantly superior due to two main reasons. Firstly, three n-bit random numbers $\{N_{T_1}, N_{T_2}, N_{T_3}\}$ take part in the generation of the session keys $\{a, b\}$. Secondly, the adversary has to eavesdrop two sessions in which $c_i = c^{*}_i$ $\forall ~ i \in \{1,...,n\}$.  This last condition is due to the fact that $Z^{0} \oplus Z^{1} \neq x$ in our proposed scheme. Summarizing, the attacker has to eavesdrop two sessions in which four nonces have the same value respectively. If we assume that the attacker is active, then he can choose the challenge sent by the reader (adversary) during the rapid bit exchange phase  but he has no control over the three random numbers generated by the tag. So, the number of sessions $N$ -- from the birthday paradox -- that an adversary has to eavesdrop to conduct a successfully dictionary attack is:
\begin{eqnarray*}
  N &\simeq& 2^{\frac{3n}{2}}\cdot \sqrt{2\cdot \ln \frac{1}{1-p}}
\end{eqnarray*}

where $n$ represents the bit length of the random numbers and $p$ is the probability of listening two sessions with the same three random numbers.

\end{remark}

Finally, we present a performance comparison of the most well-known existing distance bounding protocols that attempt to offer resistance against both mafia and terrorist fraud attacks.  Table \ref{table::pcomparison} summarizes our assessment. Firstly, in Table 1 (A), we indicate if these protocols are vulnerable to any of these frauds. In case they are, we give the corresponding success probability for an adversary. That is, the probability an adversary deceives a valid reader into believing that it is communicating with a valid tag and that this tag is within a particular physical distance.  This probability is commonly appeared in the literature as the False Acceptance Ratio (FAR). More precisely,  M-FAR and T-FAR represent the false acceptance ratio against mafia and terrorist fraud attacks respectively.  The columns Mafia and Terrorist show whether the defence against each of these attacks is an objective of the protocol.

In Table 1 (B), we first indicate what are the protocols that have a mechanism for handling errors (``Error resistance'' column). This is an issue that must be addressed explicitly, since the challenge-response bits transmitted in the rapid bit exchange phase are sensitive to channel errors. Secondly, we indicate whether the protocols include privacy protection measures (``Privacy'' column). It is easy to see that the majority of the proposals put the privacy of tag holders at stake due to two main reasons: 1) tags/readers transmit in plain-text their identities; 2)  all the tags share a single secret. Thirdly, in ``Mutual Authentication'' column we indicate which protocols provide mutual authentication between the verifier (reader) and the prover (tag). Finally, we calculate the computation overhead. More precisely, we indicate the number of invocations -- ``Operations'' column -- of cryptographic primitives such as hash functions, pseudo-random functions or symmetric key encryptions. Finally, we list the required number of random numbers (nonces) for the execution of each protocol (``Nonces'' column).

\begin{table*}
  \centering
  \caption{Performance Comparison of distance bounding protocols}\label{table::pcomparison}
\begin{scriptsize}
\begin{tabular}{|c|c|c|c|c|}
  \cline{2-5}
  \multicolumn{1}{c}{} & \multicolumn{4}{|c|}{Fraud} \\ \hline
  Protocol & Mafia & M-FAR & Terrorist & T-FAR  \\ \hline
  Brands \& Chaum \cite{brands94} & Yes & $(\frac{1}{2})^n$ & No & --  \\ \hline
  Hanche \& Kuhn \cite{hancke05} & Yes & $(\frac{3}{4})^n$ & No & --  \\ \hline
  Reid et al. \cite{reid2007}$^\S$ & No & -- & No & -- \\ \hline
  Tu \& Piramuthu \cite{TuP-2007-rfidtechnology}$^\S$  & No & -- & No & --  \\ \hline
  Swiss-Knife \cite{KimAKSP-2008-icisc}$^\S$ & No & -- & No & -- \\ \hline
  Hitomi  & Yes & $(\frac{1}{2})^n$ & Yes & $(\frac{1}{2})^n$ \\
  \hline
\end{tabular}
\begin{quote}
   ~~~~~~~~~~~~~~~~~~~~~~~~~~~~~~~~~~~~~~~~~~~~~~~~~ \textbf{Table 1 (A)}
\end{quote}

\begin{tabular}{|c|c|c|c|c|c|}
  \multicolumn{1}{c}{}&   \multicolumn{5}{c}{} \\ \hline
  Protocol & Err. resis. & Privacy & M. Authen. & $\sharp$ Op. & $\sharp$ Nonces   \\ \hline
  Brands \& Chaum \cite{brands94} &  No &  -- & No & 2 & 1 \\ \hline
  Hanche \& Kuhn \cite{hancke05} & Yes &  -- & No & 1 & 1 \\ \hline
  Reid et al. \cite{reid2007}$^\S$ & &  No & No & 2 & 1 \\ \hline
  Tu \& Piramuthu \cite{TuP-2007-rfidtechnology}$^\S$  &  Yes &  No & No & 5 & 1 \\ \hline
  Swiss-Knife \cite{KimAKSP-2008-icisc}$^\S$ & Yes &  No & No & 3(2$^\dag$) & 1\\ \hline
  Hitomi  & Yes &  Yes & Yes & 6(5$^\dag$) & 3$^\ddag$\\
  \hline
\end{tabular}
\begin{quote}
   ~~~~~~~~~~~~~~~~~~~~~~~~~~~~~~~~~~~~~~~~~~~~~~~~~ \textbf{Table 1 (B)}
\end{quote}

\begin{quote}
    $^\S$  The protocol is vulnerable to a full disclosure attack which wrecks all the security properties claimed (see Sections \ref{sec::secattack} and \ref{sec::experiments}). \\
    $^\dag$ Reader authentication is not a protocol requirement.   \\
    $^\ddag$ This is the first protocol that considers and offers resistance to dictionary attacks.
\end{quote}

\end{scriptsize}
\end{table*}

\subsection{The Threshold}
In our protocol, we perform a rapid bit exchange composed of $n$
challenge-response rounds.  At the end of the rapid bit exchange we
count the number of misses (errors) $\err$ between the transmitted and
received challenges and responses.  More precisely, we consider that a
transmitted challenge $c_{i}$ might be different from the received
challenge $c'_{i}$ and similarly a transmitted response $r_{i}$ might
be different from a received response $r'_{i}$. These mismatches
between $c_{i}$, $c'_{i}$ or $r_{i}$, $r'_{i}$ might be caused either
due to noise in the communication channel or due to the fact that the
tag that tries to be authenticated is not legitimate (an
adversary).

We need to tradeoff the possibility of authenticating an adversary and
rejecting a legitimate tag due to some legitimate errors caused by the
noise in the communication channel.
For this purpose, we introduce a threshold $\thr$ such that we shall
authenticate the communicating party if and only if $\err < \thr$.

To fully specify the problem, we assume that if we wrongly
authenticate an attacker $A$, then we suffer loss $\loss = \LA$. On
the other hand, if we reject a legitimate tag $T$, then we suffer a
loss $\loss = \LT$. In other cases, we suffer no loss.\footnote{These
  losses are subjectively set to application-dependent
  values. Clearly, for cases where falsely authenticating an attacker
  the impact is severe, $\LA$ must be much greater than $\LT$.} We
wish to minimise the expected loss $\E \loss$. The expected loss given
that the communicating party is an attacker $A$ or the tag $T$, is
given respectively by:
\begin{equation}
  \E (\loss | A) =
  \Pr(\err < \thr | A) \cdot \LA
  + \Pr(\err \geq \thr | A) \cdot 0 \nonumber
\end{equation}
\begin{equation}
  \E (\loss | T) =
  \Pr(\err < \thr | T) \cdot 0
  + \Pr(\err \geq \thr | T) \cdot \LT. \nonumber
\end{equation}
The expected loss is in any case bounded by the maximum loss:
\begin{equation}
  \E \loss \leq \max \left\{ \E(\loss | A), \E(\loss | T)\right\}.
  \label{eq:maximum-loss}
\end{equation}
If the attacker $A$ has an error rate that is at least $\pa$, then we can
analytically express a bound on the probability of falsely
authenticating him via the binomial formula. However, a simpler
expression is given by the following inequality, which holds for $\pa >
\thr/n$, by noting that $\E(\err | A) = n \pa$:
\begin{eqnarray*}
  \Pr(\err < \thr | A)
  &<&
  \exp\left(-2n(\pa - \frac{\thr}{n})^2\right)
  = \\
  &=& \exp\left(-\frac{2}{n}(n\pa - \thr)^2\right) 
\end{eqnarray*}
This bound follows from the well-known Hoeffding inequality (Equation
~\eqref{eq:hoeffding} in Appendix).  Similarly, we can bound the probability of
falsely rejecting a tag $T$ with error rate $\pt$ by:
\begin{eqnarray*}
  \Pr(\err \geq \thr | T)
  &\leq&
  \exp\left(-2n(\pt - \frac{\thr}{n})^2\right)
  =\\
  &=&\exp\left(-\frac{2}{n}(n\pt - \thr)^2\right) 
\end{eqnarray*}
for $\pt < \thr / n$.

Henceforth, we always consider the case $\pt < \thr / n < \pa$.
Equating the two terms in the max operator of \eqref{eq:maximum-loss}
we obtain, after some elementary manipulations, a value of $\thr$ such
that they are approximately equal:
\begin{align}
  \thr
  =
  \frac{2n(\pt^2-\pa^2) - \log \frac{\LT}{\LA}}{4(\pt -\pa)}.
  \label{eq:threshold}
\end{align}
This can be used as a nearly optimal value for the threshold $\tau$, given
the bounds $\pt, \pa$ on the error probability of the user and attacker
respectively, and the desired losses $\LA, \LT$.

\subsubsection{Choice of the Threshold for Specific Protocols}

In this section we  calculate the values of $\pa$ and $\pt$ for Kim et al.'s (Swiss-Knife) protocol \cite{KimAKSP-2008-icisc}. These are also valid for the proposed protocol (Hitomi) and can be used to select an appropriate value of the threshold $\thr$ for a given number ($n$) of rapid bit exchange challenge-response rounds.

In Kim et al.'s distance bounding protocol the condition that aborts  the authentication of the prover (tag $T$) is the following:
\begin{equation}
err_{c}+err_{r}+err_{t} \geq \thr, \nonumber
\end{equation}
where $\forall i \in [1,n]$:
\begin{align}
err_{c}:=& \#\{i : c_{i}\neq c'_{i}\} \\
err_{c}:=& \#\{i : c_{i}= c'_{i} \AND  r_{i} \neq Z_{i}^{{c_{i}}} \}\\
err_{t} :=& \# \{i: c_{i}=c'_{i}   \AND \Delta t_{i} >t_{max}\}
\end{align}

The optimal strategy for the attacker cannot include delaying to send his responses.
To see this imagine that the attacker can choose to either wait in order to obtain the correct challenge-response pair or to guess. If he does the former then the number of errors $err_{t}$ increases by one always. If however he guesses, then there is always a non-zero probability that the number of errors will not increase.
Thus, while the third error $err_{t}$ is useful for detecting classic relay attacks, it cannot be part of the attacker's strategy for any of the other attacks. That means that if the attacker has the choice to mix the delaying strategy with a guessing strategy, it is always better to choose the guessing strategy. For this reason, the error $err_{t}$ is disregarded in the analysis.

Thus, in order to minimise the expected loss, we only need to consider the following condition:

\begin{align}
err_{c}+err_{r} \geq \thr
\end{align}

Let us assume that $c_{i}$ denotes the challenge sent by the reader in
the $i$-${th}$ round of the rapid bit exchange phase. We denote by
$c'_{i}$ the challenge that the legitimate tag $T$ received. We assume
due to noise, that $c_{i} \neq c'_{i}$ can occur with probability $\noise
\in [0, 1/2]$.  Finally, we use $c''_{i}$ to denote the challenge
received by the attacker $A$, which may again differ from $c_{i}$ with
probability $\noise$.





For the attacker $A$, the probability of making an error $\err$ is:
\begin{align}
 \Pr(\err | A) &=
 \Pr (\err | c \neq c') \Pr(c \neq c') +   \Pr (\err | c = c') \Pr(c = c')   \nonumber \\
&= 1\cdot \frac{1}{2} +  \Pr (\err | c = c')\cdot \frac {1}{2}
\label{eq:perrorAttacker1}
\end{align}
It holds that:
\begin{align}
  \Pr(\err | c=c')
  =& \Pr(c''=c | c=c' )  \Pr(\err | c'' = c \AND  c=c')   \nonumber\\
  + & \Pr(c'' \neq c | c=c' )  \Pr(\err | c'' \neq c \AND  c=c')  \nonumber \\
  =&(1 - \noise) \cdot \noise + \noise \cdot \frac {1} {2} = \frac {3
    \noise} {2} - w^{2}
 \label{eq:perrorAttacker2}
\end{align}
Thus, equation \eqref{eq:perrorAttacker1} using
\eqref{eq:perrorAttacker2} gives us:
\begin{equation}
 \Pr(\err | A) = \frac{1}{2} +  \frac {3 \noise} {4} - \frac {\noise^{2}} {2}  \geq \frac {\noise +1} {2}
 \label{eq:perrorAttacker}
\end{equation}
assuming that $\noise \leq \frac {1} {2}$ and $\noise \neq 0$.
For the legitimate tag $T$ the probability of making an error $\err$
is given by:
\begin{align}
 \Pr (\err | T) = \nonumber \\
\Pr (\err | c'' =c') \Pr(c=c') + \Pr (\err | c'' \neq c') \Pr(c\neq c') = \nonumber \\
\noise  (1 - \noise) + \noise = 2 \noise - \noise ^{2} \leq 2 \noise
\label{eq:perrorTag}
\end{align}
From, equations \eqref{eq:perrorAttacker} and \eqref{eq:perrorTag} we get:
\begin{equation}
\pa=\frac {\omega + 1} {2}  \mbox{ and } \pt=2\noise
\end{equation}
By substituting the values of $\pa$, $\pt$ in  equation \eqref{eq:threshold} we get:
\begin{equation}
\thr = \frac {n (5 \noise +1)}{4} - \frac {log\rho}{6\noise -2},
\end{equation}
where $\rho=\frac {\LT} {\LA}$.

\begin{figure}
\centering
  \includegraphics[width=8cm]{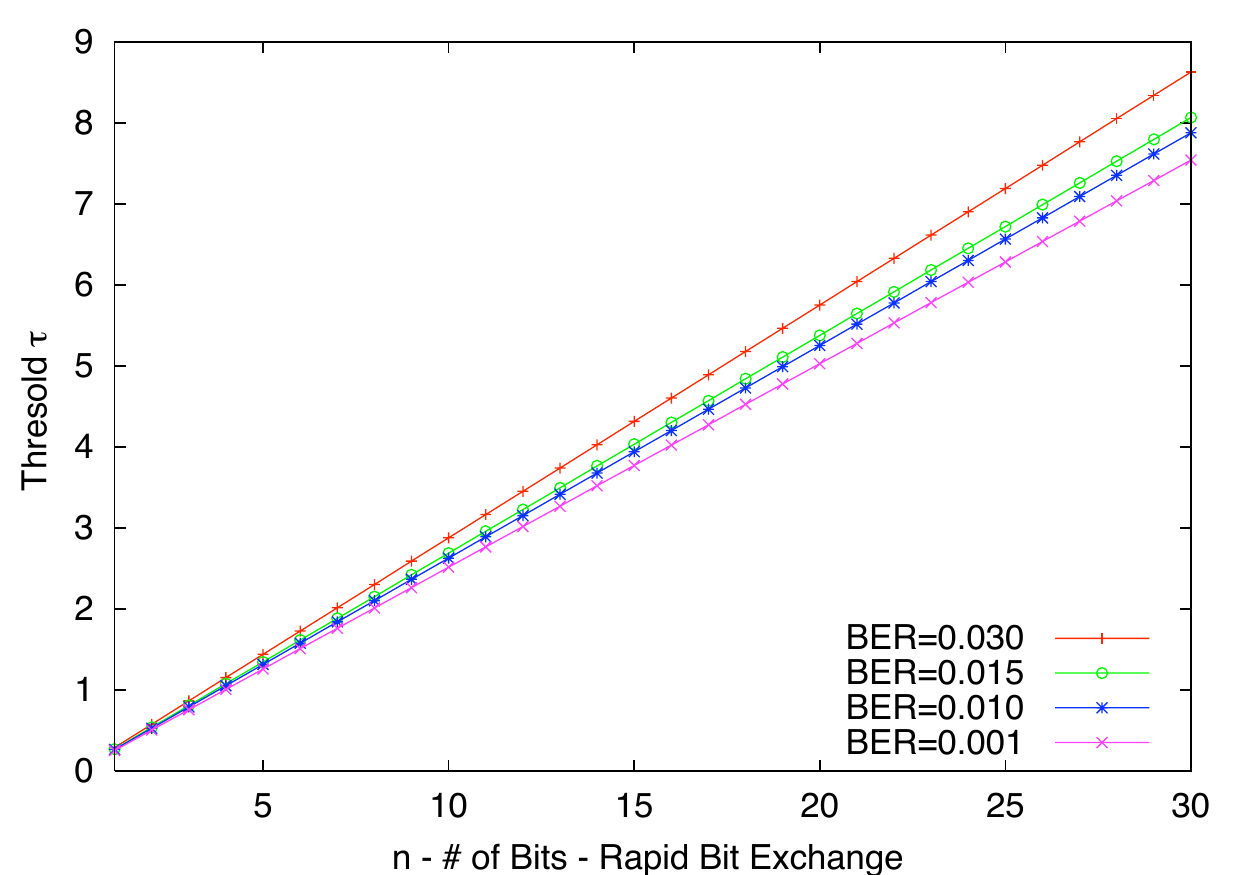}
  \caption{The Threshold $\thr$  for various values of the probability of noise $\omega$ (BER) when $\rho=\frac {\LT} {\LA}=1$ }\label{fig::figu1}
\end{figure}

\begin{figure}
\centering
  \includegraphics[width=8cm]{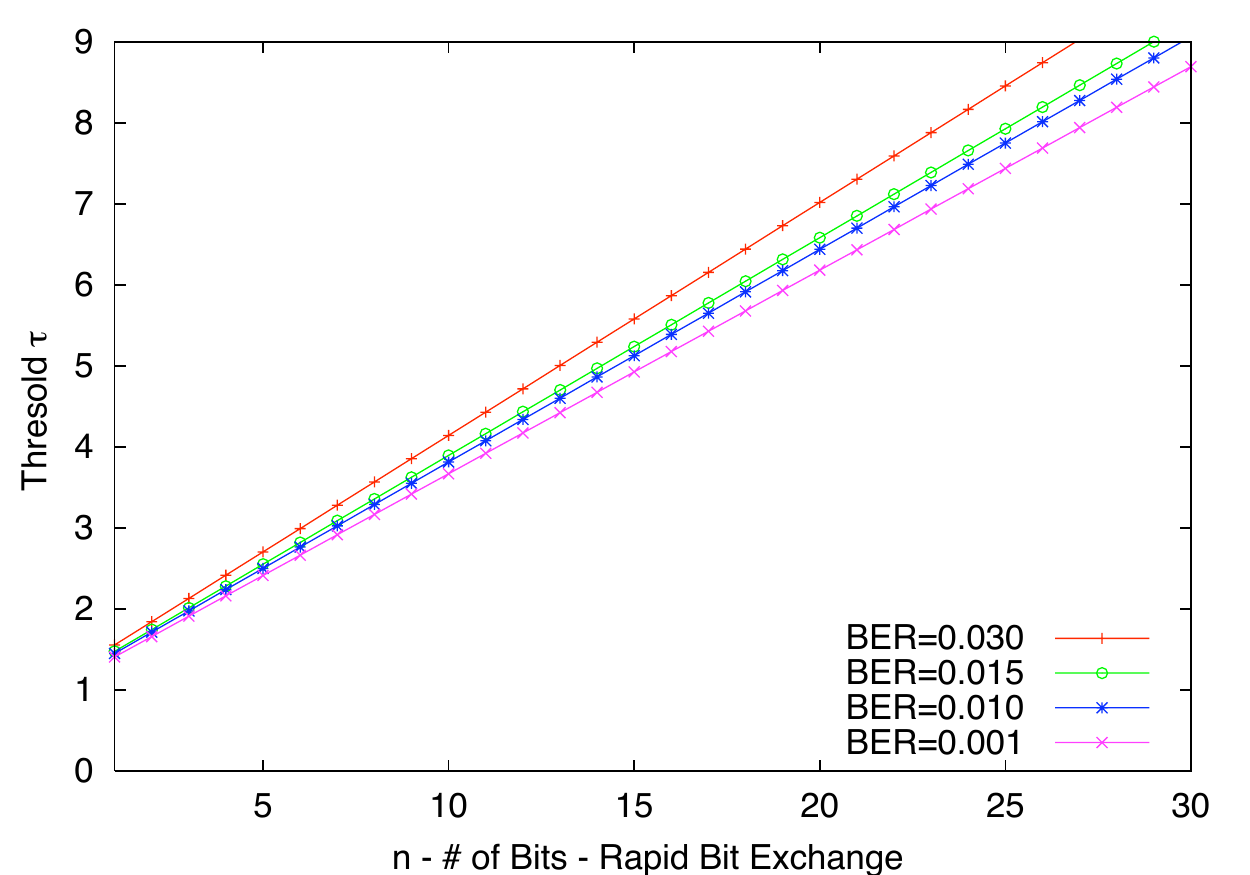}
  \caption{The Threshold $\thr$  for various values of the probability of noise $\omega$ (BER) when $\rho=\frac {\LT} {\LA}=10$ }\label{fig::figu2}
\end{figure}

Fig. \ref{fig::figu1} and \ref{fig::figu2} depict how the value of the threshold $\thr$ changes as we increase the number of bits ($n$) exchanged during the rapid bit exchange for various values of the probability of noise $\noise$ (BER= $\{0.030, 0.015,0.010,0.01\}$). Fig. \ref{fig::figu1} depict the results when $\rho=\frac {\LT} {\LA}=1$, while figure \ref{fig::figu2} depict the results when $\rho=\frac {\LT} {\LA}=10$. In both cases it is obvious that the threshold $\thr$ reduces as the noise $\noise$ reduces and increases as the numbers of bits $n$ increases. Additionally, when the ratio $\rho$ increases then the threshold $\thr$ also increases.

Kim et al. \cite{KimAKSP-2008-icisc} introduced the use of a threshold $\tau$ in a distance
bounding protocol that can be used to avoid the failure of
authenticating legitimate tags considering that some legitimate errors
might be caused due to the noise in the communication channel.
However, to the best of our knowledge, this is the first time that a
detailed analysis of the threshold $\tau$ is provided taking into
consideration the probability of having errors due to the noise in the
communication channel.

\section{Conclusions}

Physical location is an important  security variable. Several techniques can be employed to measure the distance between a prover and a verifier. However, signal strength is an unreliable indicator of distance, since it can be manipulated. For this reason, delay-time based measurements such as those employed in distance bounding protocols, are more appropriate.  Nevertheless, the vast majority of these protocols focus on thwarting mafia fraud attacks. This paper is concerned with terrorist attacks, whose security considerations have not received a lot of attention. In fact, none of the currently existing proposals are completely secure against terrorist fraud attacks.  In 2008, Kim et al. \cite{KimAKSP-2008-icisc}  proposed a distance bounding protocol (Swiss-Knife protocol), which was claimed to cover all security objectives expected from an RFID system. However,  we assume that Swiss-Knife protocol is vulnerable to a passive attack, which may lead to the full disclosure of the secret key shared between the reader and the tag.  Consequently,  all security objectives are compromised. Furthermore, the described attack is also applicable to the protocols proposed by Reid et al.  \cite{reid2007} and Tu and Piramuthu \cite{TuP-2007-rfidtechnology}, since the Swiss-Knife protocol is based on these schemes.  Additionally, we provide  a set of guidelines that should be followed by protocol designers in order to design secure and efficient distance bounding protocols suitable for devices with constrained resources.  Finally, a new protocol, named Hitomi, conforming to the proposed guidelines is presented and analyzed from the point of view of security and performance.  The existence of more efficient protocols that can offer the same level of security is an open problem.

\appendix

\begin{lemma}[Hoeffding]
  For any sequence of random variables $X_1, \ldots, X_n$ such that
  $X_i \in [a_i, b_i]$, with  $\mu_i \defn \E X_i$:
  \begin{equation}
    \Pr\left(
      \sum_{i=1}^n X_i \geq \sum_{i=1}^n \mu_i + n t
    \right)
    \leq
    \exp\left(
      -2\frac{n^2t^2}{\sum_{i=1}^n b_i - a_i}
    \right).
    \label{eq:hoeffding}
  \end{equation}
  \label{lem:hoeffding}
\end{lemma}

\end{document}